\documentclass[aps,onecolumn,pre]{revtex4}
\usepackage{latexsym, verbatim}
\usepackage{amsmath}
\usepackage{mathrsfs}
\usepackage{amsthm}
\usepackage{amssymb}
\usepackage{dsfont}
\usepackage{fancybox}
\usepackage{epsfig}
\usepackage{bm}
\usepackage{fancyhdr}
\usepackage{subfigure}
\usepackage{graphicx}
\usepackage{color}

\begin{document}
 
\title{\bf The Drastic Role of Beyond Nearest-Neighbor Interactions on 
Two-Dimensional Dynamical Lattices: A Case Example}

\author{P.G. Kevrekidis}
\affiliation{Department of Mathematics and Statistics, University of 
Massachusetts,
Amherst MA 01003-4515, USA}
%\date{\today}

%\maketitle

\begin{abstract}
In the present work, we highlight the significant effect that 
the simplest beyond nearest neighbor interactions can have on 
two-dimensional dynamical lattices. To do so, we select as our 
case example the 
closest further neighbor, namely the diagonal one, and a prototypical
nonlinear lattice, the discrete nonlinear Schr{\"o}dinger 
equation. Varying solely the strength of this extra neighbor
interaction, we see examples of
(a) destabilization of states that start out as stable in
the nearest neighbor limit; (b) stabilization of states that
start out as unstable in that limit; (c) bifurcation of novel 
states that do not exist in the nearest neighbor case. These 
dramatic changes are first theoretically highlighted through an
analysis of a reduction of the problem to a few excited sites
and the associated set of conditions that govern their existence
and their dynamical stability. Then, they are corroborated
numerically through fixed point computations, spectral analysis
and nonlinear dynamical evolution simulations.
\end{abstract}

\maketitle

%\vspace{-10mm}

%\textit{Introduction}.

\section{Introduction}

The broad theme of intrinsic localized modes has attracted a significant
volume of attention from a wide range of communities over the past two
decades~\cite{review}. Such interest stems from their emergence not
only in the modeling and computation, but perhaps importantly in 
experiments of a wide array of themes. These include, but are not
limited to, arrays of nonlinear-optical waveguides \cite{moti}, 
Bose-Einstein condensates (BECs) in periodic potentials \cite{ober},
micromechanical cantilever arrays~\cite{sievers}, 
Josephson-junction ladders \cite{alex},
granular crystals of beads interacting through Hertzian contacts~\cite{theo10}, layered antiferromagnetic crystals~\cite{lars3},
halide-bridged transition metal complexes~\cite{swanson},
 and dynamical models of the DNA double strand \cite{Peybi}.

Among the many models of discrete systems (i.e., nonlinear dynamical 
lattices) that have been proposed for these physical settings, one
that has been a central point of attention is the so-called
discrete nonlinear Schr{\"o}dinger (DNLS) equation~\cite{book}. 
Part of the intrigue in this model lies in its apparent simplicity
(since it incorporates solely the prototypical characteristics
of interest, namely nonlinearity and a discrete form of dispersion),
yet its considerable wealth of nonlinear wave solutions and
phenomena. Furthermore, its relevance as a suitable approximation
of optical (in waveguides)~\cite{dnc,moti} and atomic systems (in optical 
lattices)~\cite{emergent} and the particularly simple form of breathers
due to its separability of space and time variables in the standing
waves of the DNLS only add to its appeal.

One of the many themes that have been considered in the context
of the DNLS equation is that of long(er) range interactions. In particular,
the interest in this theme concerns the effects of potential inclusion
of interactions beyond those of the purely nearest neighbors of the
standard DNLS.  In that context, numerous interesting possibilities
have been found to arise. For example, it has been shown that 
interaction strengths with sufficiently slow decay (over space)
can give rise to bistability of fundamental solitary waves
(centered on a single lattice site)~\cite{nonlocalBistability}.
This, in turn, may play a role in soliton 
switching~\cite{switchingBistable}. On the other hand, such
interactions may be relevant for energy and charge transport
in biomolecules~\cite{EnergyChargeTransportBiomolecules} and
polymers~ \cite{nextnearestPolymer}, as well as in waveguide
arrays. In the latter setting, one possibility for their relevance
is near the so-called zero-dispersion point~\cite{borisus}. But,
arguably, a more relevant example is that of zigzag-shaped waveguide
arrays which have not only been theoretically proposed~\cite{Cyprus},
but also experimentally implemented and demonstrated to be valuable
for promoting localization instead of diffraction even in the linear
regime~\cite{alexs}. As an aside, it is relevant to note that
quantum variants of the nonlocal DNLS equation have been studied by means
of the Bethe ansatz~\cite{quantum}. Furthermore, the role 
of nonlocal interactions in BECs
has also been proposed to be critical in stabilizing rather unusual
spatially periodic states such as the 3-site period waveforms
identified in a discrete, long-range DNLS-type model proposed
for $^{52}$Cr,
in the presence of an optical lattice~\cite{borispfau}.
Although, a general formulation can be provided~\cite{book,pla1d} for the
existence and stability of DNLS standing 
waves from the well-known anti-continuum
limit (of vanishing coupling between adjacent sites~\cite{macaub}),
the number of studies that tackle nonlinear waves in higher
dimensional longer range settings is very limited, to our knowledge,
and chiefly restricted to fundamental solutions; 
see e.g.~\cite{2DcompetingShortLongRange}. Naturally, with the computational
and analytical tools that are presently available, it is relevant to 
seek a deeper understanding of the role of longer range
interactions in these and other wave systems, and some recent studies
have been aiming in that direction, such as the Klein-Gordon chain analysis
of~\cite{vassilis}.

Such an improved understanding in a special and suitably tailored,
but also as we will demonstrate quite rich in its phenomenology, 
case example is the scope of the present work. In particular, we
will focus on the setting where solely nearest neighbor and
next-nearest-neighbor (that is, diagonal in our two-dimensional
lattice) interactions are present.
In this realm, we will demonstrate that the beyond-nearest-neighbor
interactions are a powerful controller of both the existence and
also of the stability of solutions and consequently of the
dynamical evolution of the system. More specifically, we will 
demonstrate that solutions (such as the discrete vortices of topological
charge $S=1$), which are robust enough that they can be observed in
photonic crystal experiments~\cite{experis1}, can be destabilized
even by arbitrarily small beyond-nearest-neighbor interaction
(of suitable sign). Moreover, we will show that other states
which are unstable in the standard case (such as the vortices
of topological charge $S=2$; see for theory~\cite{peli_2d,book}
and for recent experiments~\cite{zhig}) will be stabilized
when the next-nearest neighbor effect is sufficiently strong.
Moreover, we will identify special limits (such as the degenerate
unit diagonal neighbor limit) whereby even solitary wave (non-vortex)
solutions will change (or exchange) their stability. This gives rise
to unusual bifurcation events, such as a double pitchfork scenario that
we identify in what follows, as well as to the emergence of novel
and previously uncovered branches that solely exist because of the
strength of the beyond-nearest-neighbor term.

We should add a note in connection to the experimental 
implementation of waveguide arrays. The typical scenario of relevance
therein involves a next-nearest-neighbor interaction which is weaker than
the nearest-neighbor one. However, as Figs. 1 and 2 of Ref.~\cite{Cyprus} 
show, it is
certainly possible in 1d zigzag settings to create a
next-nearest-neighbor interaction
which is stronger than the nearest-neighbor 
one. This can be done systematically by
simply modifying the angle of the zigzag lattice. On the other hand, in 
2d admittedly this possibility is harder to realize. However, as is 
discussed in~\cite{pra_alex,opex_alex},
%in the newly added references [Phys. Rev. A 77, 043804 (2008)
%and Opt. Express {\bf 15}, 1579 (2007)], 
in the latter setting as well, it is possible
to tune the interactions in femtosecond laser written waveguides with 
elliptical shape, by tilting the elliptical waveguides. While we are
not aware presently 
of a 2d setting where such tilting (or other technique)
has been able to produce a dominant next-nearest-neighbor interaction, 
nevertheless that regime is certainly
of theoretical interest and, given the rapid progress of corresponding 
experimental technology, may soon become experimentally relevant as well,
hence it is also considered here.

Our approach in what follows will be two-pronged and will be based
on a theoretical analysis of this setting from the anti-continuum
limit and an accompanying set of continuation/bifurcation and
dynamical evolution numerical
computations from that limit. This will enable us to obtain systematic
information about the existence of solutions, and also about their
linear stability and to test these predictions against the corresponding
numerical computations. Finally, the results will also be corroborated
with direct numerical simulations to illustrate the stable or unstable
(as appropriate) evolution of our identified waveforms. 
After presenting the model and theoretical setup in 
section II, our analytical and numerical results and their comparison
will be given in section III.
Finally, in section IV, we offer a summary of the distinguishing features
induced by the beyond nearest neighbor interactions in this system
and a number of associated conclusions and potential future directions
for the extension of the present study.

\section{Model and Theoretical Setup}

The DNLS equation on the two-dimensional square lattice 
of interest herein has a standard form \cite{book},
\begin{equation}
i\frac{d}{dz}\phi _{m,n}+ \epsilon {\cal L}\phi _{m,n}+\left\vert \phi
_{m,n}\right\vert ^{2} \phi_{m,n}=0,  \label{NLS}
\end{equation}
where $\epsilon$ is the coupling constant, and the linear 
operator ${\cal L}$ typically assumes the form of the
discrete Laplacian $\Delta _{2}$. The latter is
defined according to $\Delta _{2}\phi _{m,n}=\phi _{m+1,n}+\phi
_{m,n+1}+\phi _{m,n-1}+\phi _{m-1,n}-4\phi _{m,n}$. 
However, for our more general case, we will assume that 
\begin{eqnarray}
{\cal L} \phi_{m,n}= \sum_{(m_1,n_1) \in NN} \phi_{m_1,n_1} + 
k \sum_{(m_2,n_2) \in NNN} 
\phi_{m_2,n_2}. 
\label{laplace}
\end{eqnarray}
That is we will consider both the effect of nearest neighbors (whose set
is denoted by NN)
and that of the next nearest neighbors (denoted by NNN).
Notice that in the above, motivated by the optical waveguides problem,
we use the optical notation where the evolution variable (propagation 
distance)
is
denoted by $z$. Also, in the expression of Eq.~(\ref{laplace}),
the onsite term $\propto \phi_{m,n}$ has been suppressed.

Looking for
stationary solutions, in the customary form $\phi _{m,n}=\exp (i\Lambda z)u_{m,n}$, Eq. (\ref{NLS}) leads to the time-independent equation:
\begin{equation}
-\Lambda u_{m,n}+ \epsilon {\cal L} u_{m,n}+ |u_{m,n}|^{2}u_{m,n}=0.  \label{standing}
\end{equation}
%We obtained 
Without loss of generality, we can rescale $\Lambda=1$.
Furthermore, to somewhat simplify notation, we will also
use the vector formulation involving ${\bf l}=(m,n)$.

Our analysis will take advantage of the well-established anti-continuum
limit~\cite{macaub}, in order to develop a perturbative analysis 
from there. In particular, in that limit (of uncoupled sites), the
energy of the decoupled oscillators assumes the form:
\begin{eqnarray}
E_0(u)=\sum_{\bf l} |u_{\bf l}|^2 - \frac{1}{2} |u_{\bf l}|^4.
\label{en0}
\end{eqnarray}
Now, introducing the coupling adds a term to the energy
with 
\begin{eqnarray}
E_1(u)=-\frac{1}{2} \epsilon \sum_{{\bf l}{\bf l'}} J_{{\bf l} {\bf l'}} \left(u_{\bf l}^{\star} u_{\bf l'} + u_{\bf l} u_{\bf l'}^{\star}
\right)
\label{en1}
\end{eqnarray}
where the $1/2$ prefactor is intended to avoid double-counting.
In the setting described above, the kernel of interaction 
is non-vanishing only for $|{\bf l}-{\bf l'}|=1$ (NN), in which case
$J_{{\bf l},{\bf l'}}=1$ and for $|{\bf l}-{\bf l'}|=\sqrt{2}$ (NNN),
in which case $J_{{\bf l},{\bf l'}}=k$.

The general persistence conditions~\cite{kapit,sand,book} 
of solutions from the anti-continuum
limit (which can be straightforwardly written as $u_{\bf l}=e^{i \theta_{\bf l}}$)
demand that the unperturbed wave corresponds to an extremum of
the perturbed energy for the solution to persist. This
{\it necessary condition}
suggests that the gradient of $E_1(u)$ evaluated at 
a solution with multiple excited sites $u_{\bf l}=e^{i \theta_{\bf l}}$
should vanish. Direct calculation of this yields for each site
${\bf l}$ the solvability condition
%\begin{eqnarray}
\begin{eqnarray}
\sum_{{\bf l'} \neq {\bf l}} J_{{\bf l}{\bf l'}} \sin(\theta_{\bf l}-\theta_{\bf l'})=0.
\label{en2}
\end{eqnarray}

Moreover, in order for the solution to be stable, the corresponding extremum
has to be a minimum of the effective energy $E_1$~\cite{sand,book} (see also
~\cite{peli_2d}). More specifically,
the eigenvalues $\gamma_j$ 
of the Hessian of $E_1$ (evaluated at the above solution
of the conditions of Eq.~(\ref{en2})), are in fact connected to  
eigenvalues $\lambda$ (bifurcating from zero, when $\epsilon$ 
becomes non-vanishing) of the original lattice dynamical problem.
The connection is given by $\lambda_j^2= 2 \epsilon \gamma_j$,
to leading order~\cite{book,pla1d,peli_2d}. 

We will apply these considerations
predominantly for the case of 4-site squares with next-nearest-neighbor
interactions, where we can extract specific analytical conclusions from
the corresponding algebraic persistence and the above stability conditions.
Nevertheless, we will also briefly mention the interest in potential
generalizations for the 8 site contours containing e.g. 
 ((-1,-1), (-1,0), (-1,1), (0,1), (1,1), (1,0), (1,-1), (0,-1)).

The theoretical results will also be corroborated by means of
full numerical solutions.
Exact solutions of Eq. (\ref{standing}) will be obtained by means of a 
Newton method.
Upon generating such stationary solutions, their stability 
is examined through 
spectral stability analysis. 
To this aim, a perturbed expression of the form
\begin{eqnarray}
\phi _{m,n}&=&\exp (i z)u_{m,n}  
%\nonumber \\
+\delta \exp (i z)[a_{m,n}\exp(\lambda t) 
+b_{m,n}\exp (\lambda^{\star} t)],  
\label{perturbed}
\end{eqnarray}
is substituted into Eq. (\ref{NLS}). Here $u_{m,n}$ is the unperturbed stationary solution, $\delta$ is an infinitesimal amplitude of the perturbation;
$\lambda$ denotes the corresponding 
eigenvalues (which are real or complex in the case of 
instability). This leads to the following linear equation for the 
perturbation eigenmode,
\begin{equation}
i \lambda \left(
\begin{array}{c}
a_{k} \\
b_{k}^{\star }
\end{array}
\right) ={\bf M} \left(
\begin{array}{c}
a_{k} \\
b_{k}^{\star }
\end{array}
\right) ,\newline
\label{omega}
\end{equation}
where ${\bf M}$ is the Jacobian matrix,
\[
{\bf M}=\left(
\begin{array}{cc}
\partial F_{k}/\partial u_{j} & \partial F_{k}/\partial u_{j}^{\ast } \\
-\partial F_{k}^{\ast }/\partial u_{j} & -\partial F_{k}^{\ast
}/\partial u_{j}^{\ast }
\end{array}
\right) \newline
,
\]
and $F_{i}$ denotes the left hand side of Eq.~(\ref{standing}) and 
the string indices $\{i,j,k\}=m+(l-1)n$, $l=1,2,..,N$, map the $N\times N$ lattice into a vector of length $N^{2}$. Numerical solutions were sought for with the Dirichlet boundary conditions at the domain boundaries.
Notice that given the localized spatial nature of the considered solutions,
we expect that our numerical observations, for the range of parameter
values considered herein, are essentially insensitive to the precise
selection of  boundary conditions. 
To generate numerically exact stationary
solutions, the fixed point algorithm was iterated until convergence
(typically with a tolerance of $5 \times 10^{-8}$). 
Upon convergence, the spectral
analysis of the stationary solutions was performed. The results are
typically shown for $21 \times 21$ site lattices.
When the solutions are found to be spectrally unstable,
direct numerical simulations are performed (typically
with a fourth-order Runge-Kutta scheme), in order to
detect the dynamical evolution of the instability.

\section{Analytical Results, Numerical Results and Comparison}

We start with a consideration of the 4-site square,
arguably the simplest two-dimensional contour that 
encompasses in a fundamental manner the higher-dimensionality
of our setting. In this case, and using the relative phase
variables $\phi_1=\theta_2-\theta_1$, $\phi_2=\theta_3-\theta_2$
and $\phi_3=\theta_4-\theta_3$ (for our 4-site contour with
phase angles $\theta_{1,\dots,4}$), we can derive the
following algebraic persistence equations from Eq.~(\ref{en2})
\begin{eqnarray}
0 &=& \sin(\phi_1) + k \sin(\phi_1+\phi_2) + \sin(\phi_1 + \phi_2 + \phi_3)
\label{peq1}
\\
0 &=& \sin(\phi_2) + k \sin(\phi_2 + \phi_3) - \sin(\phi_1)
\label{peq2}
\\
0 &=& \sin(\phi_3) - k \sin(\phi_1 + \phi_2) - \sin(\phi_2).
\label{peq3}
\end{eqnarray}

It is now possible to manipulate the corresponding equations
to get the set of solutions available for the system. As an indication
of how to approach this problem, we note that adding 
Eq.~(\ref{peq1}) and Eq.~(\ref{peq3}), we obtain
$\sin(\phi_1)+\sin(\phi_3) = \sin(\phi_2) - \sin(\phi_1+\phi_2+\phi_3)$,
which upon subsequent use of double angle formulas results in the
conditions: either $\sin(\frac{\phi_1+\phi_3}{2})=0$
or $\cos(\frac{\phi_1-\phi_3}{2})=-\cos(\frac{2 \phi_2+\phi_1+\phi_3}{2})$.

Analysis of the resulting trigonometric conditions yields
the following branches of solutions.

\begin{enumerate}

\item The standard discrete vortex of topological charge 
$S=1$. This is the solution with ${\bf \theta}=(0, \pi/2, \pi, 3 \pi/2)$,
and $\phi_1=\phi_2=\phi_3=\pi/2$. This solution is 
well-known~\cite{book,peli_2d} to be
stable for the nearest neighbor model of $k=0$ with a double eigenvalue pair
(to leading order) $\lambda_{1,\dots,4}=\pm 2 \epsilon i$, a
double eigenvalue at $0$ (due to the phase or gauge invariance of
the model) and a higher order eigenvalue [that was calculated in~\cite{peli_2d}
as $\lambda_{5,6}=\pm \sqrt{32} \epsilon^{3/2} i$]. In addition to
illustrating the persistence (at least to the considered
leading order) of such a solution, we have computed the Hessian of
the perturbation energy of Eq.~(\ref{en1}) and have obtained the theoretical
predictions for the corresponding eigenvalues in the presence of the
next-nearest neighbor interactions, parametrized by $k$. We find
that $ \lambda_{1,\dots,4}=\pm 2 \sqrt{\epsilon k} i$ is a double
eigenvalue pair (to leading order), while the leading order 
prediction for the remaining four eigenvalues is $0$. Two of these
will stay at $0$ due to the above mentioned invariance, while the
other pair will bifurcate to higher order. Yet, here we would still
like to focus on the significance of the lower order nearest-neighbor
effect. It is remarkable that for these structures (which are called
super-symmetric in~\cite{peli_2d,book} because the leading order
- $\lambda \propto \sqrt{\epsilon}$ - does not contribute to their eigenvalues),
the NNN effect is, according to this prediction, 
the {\it dominant} one. In fact, we
can use an arbitrarily weak (even infinitesimally small in comparison
to NN) ``negative coupling'' to render the configuration unstable.
It should be noted that in the spirit of diffraction management
and diffraction engineering~\cite{mark1,mark2}, this is certainly
a scenario of potential physical interest. Nevertheless, we will
also examine below numerous cases where positive $k$ may have
interesting implications on nonlinear wave stability as well.

A relevant example of the corresponding branch of solutions 
is shown in Fig.~\ref{lfig1}. Numerical computations have been
performed for $\epsilon=0.001$. The top left set of panels showcases
the positive $k$ scenario (of stability), while the top right ones
the negative $k$ scenario (of instability). In fact, we observe
that the situation is even more complicated because apparently higher
order effects lead this double eigenvalue pair to be complex
(although its real part is captured almost perfectly by our analytical
prediction). The clear destabilization of the latter case is illustrated
further in the bottom plot of $k=-1.5$. The dynamical evolution 
suggests a symmetry breaking between the amplitudes 
of the 4 sites (which start out as equal) 
that subsequently leads to a nearly periodic
exchange of power between the 4 principal sites participating in the
vortical structure.

\begin{figure}[tbp]
\begin{center}
\epsfig{figure=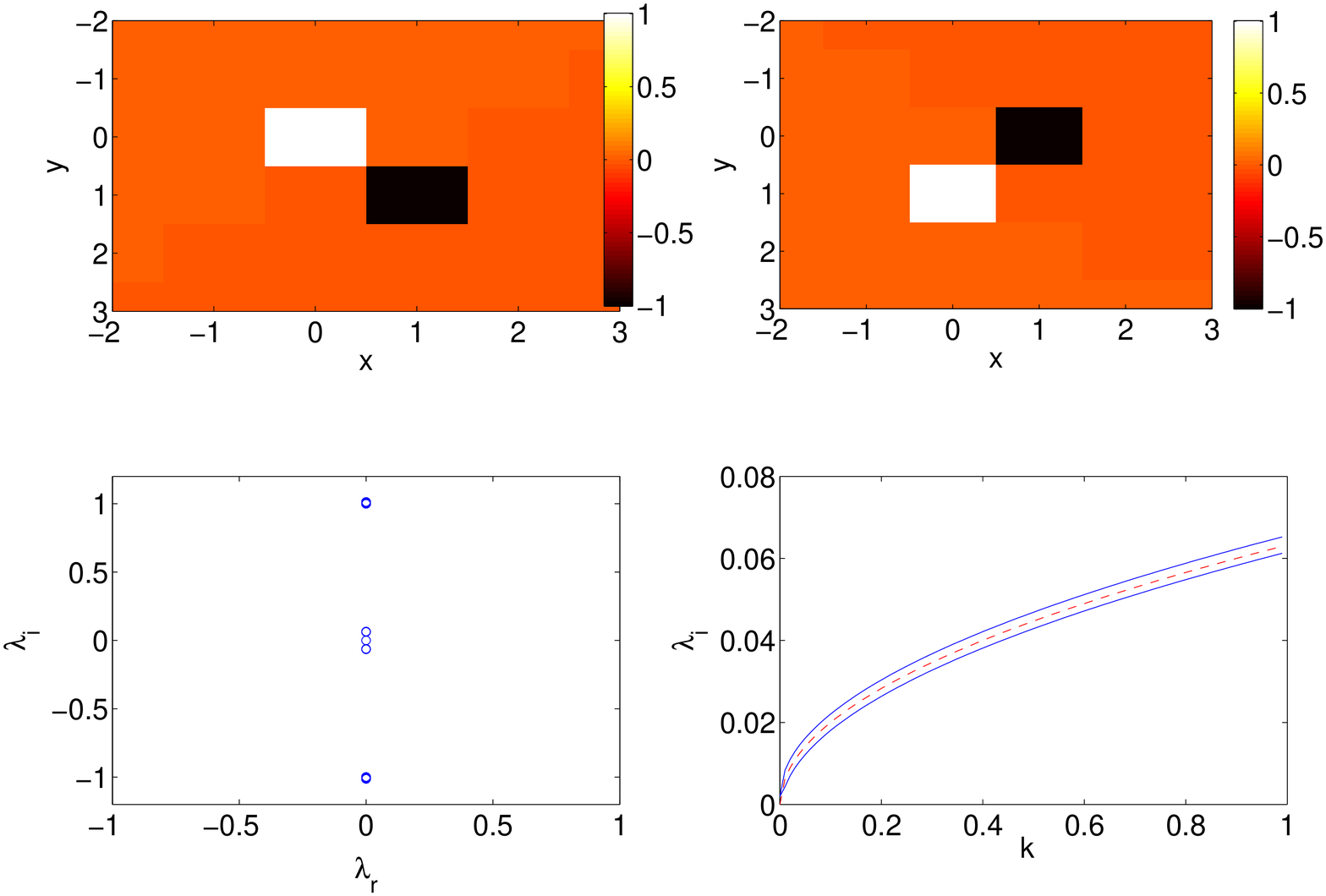,height=2in,width=3in,angle=0}
\epsfig{figure=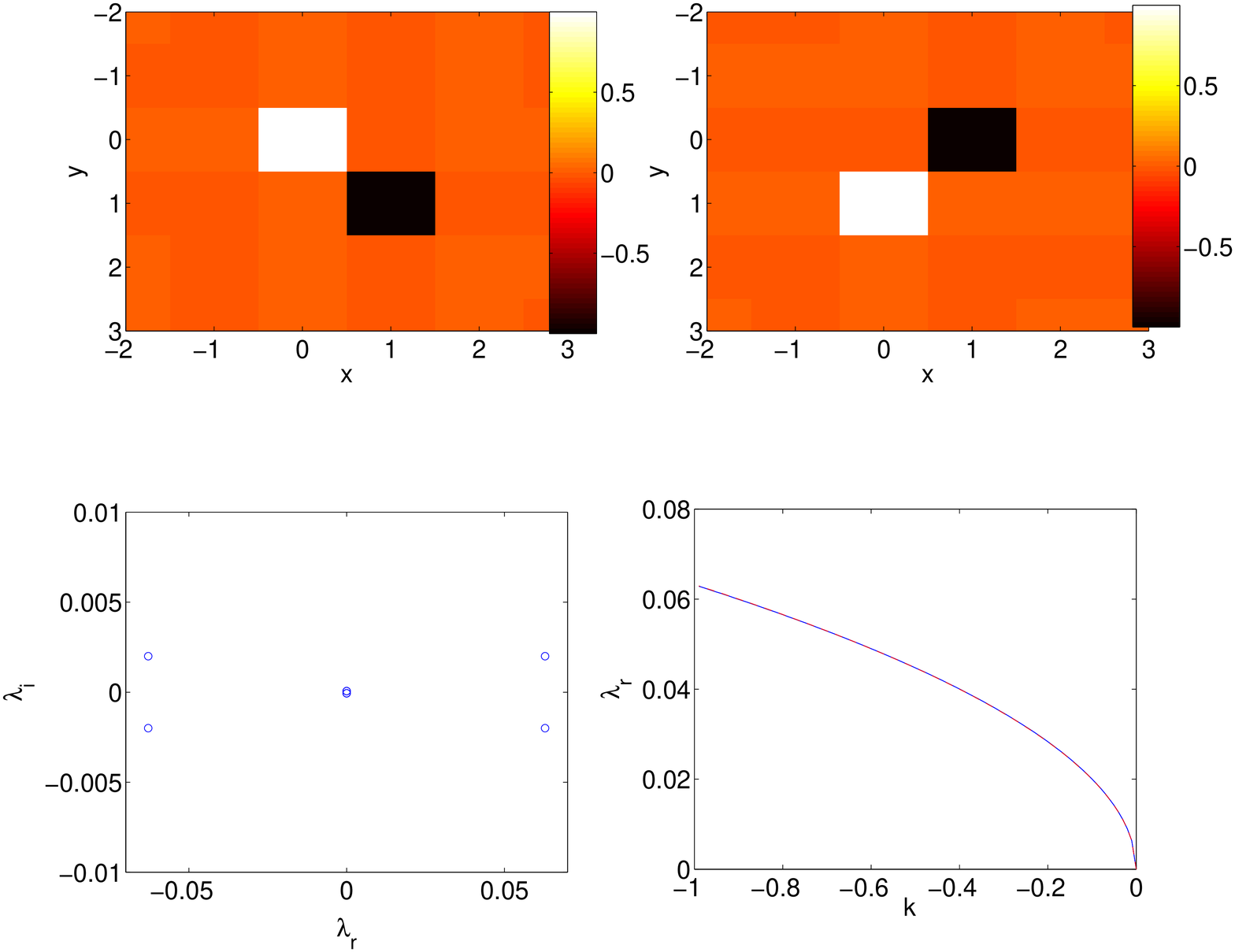,height=2in,width=3in,angle=0}
\end{center}
\begin{center}
\epsfig{figure=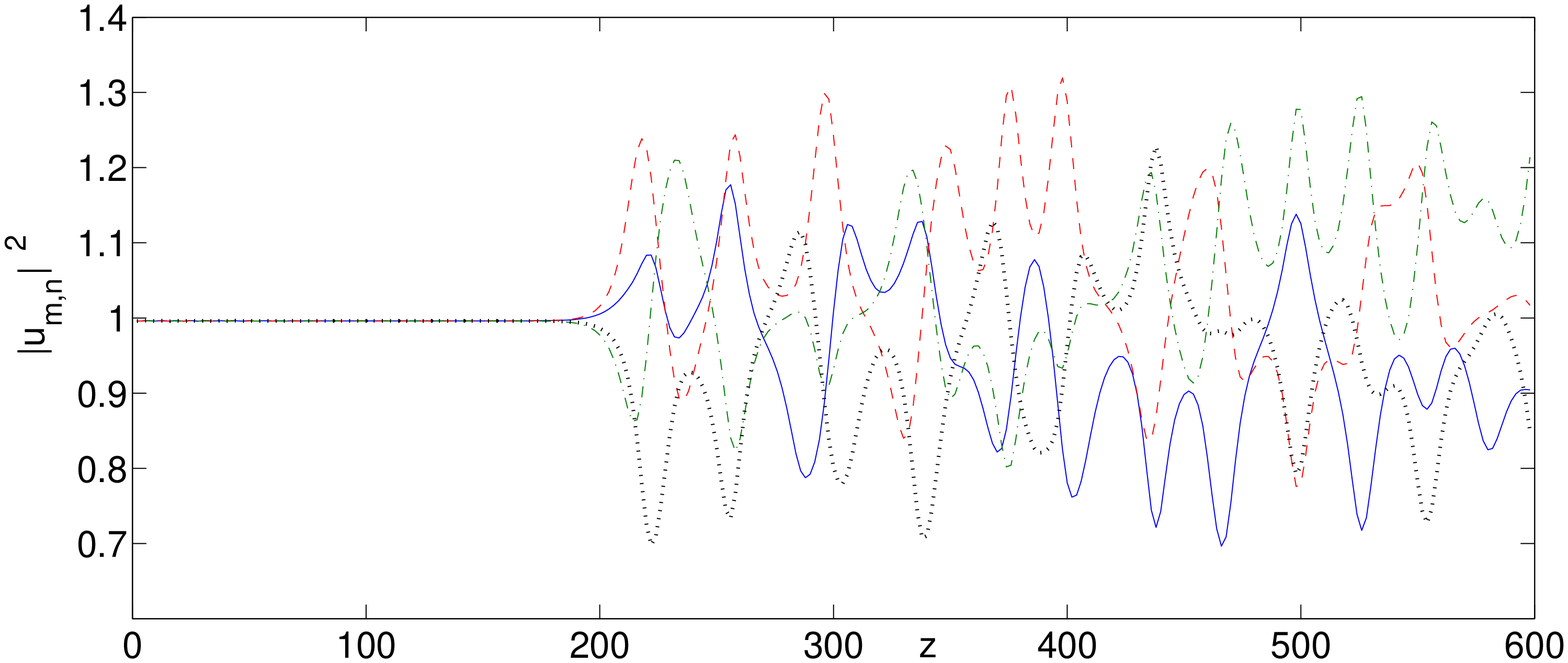,height=2in,width=5in,angle=0}
\end{center}
\caption{The top left quartet of panels shows the real and imaginary
parts of the solution, and its spectral plane $(\lambda_r,\lambda_i)$
with $\lambda=\lambda_r+i \lambda_i$ of linearization eigenvalues
-all for $k=1$-, 
as well as the
dependence of the key (theoretically predicted) eigenvalues as a function
of the next nearest neighbor strength of $k$. In the subplot that
shows the eigenvalue dependence on $k$, the
blue solid lines represent the numerical results, while the red dashed
ones the analytical prediction.
%; here the subscript r denotes the real
%part of the eigenvalue, while the subscript i the imaginary part thereof. 
Notice the very good agreement between the two.
While the left panel is for $k>0$, the right panel shows the case
of $k<0$ (the specific solutions and spectral analysis shown are
for $k=-1$). The bottom panel illustrates the evolution of the 
intensity of the 4
principal sites of such a vortex (for $k=-1.5$) 
over the propagation distance
$z$. In all of these graphs also below, the blue solid line will correspond
to the site (0,0), the red dashed line to (0,1), the green dash-dotted
to (1,0), and the black dotted to (1,1) for this 4-site contour.}
\label{lfig1}
\end{figure}

\item Another interesting solution is the so-called out-of-phase
configuration with ${\bf \theta}=(0, \pi, 0, \pi)$. In this case
all the $\phi$'s are equal to $\pi$. This configuration is also well
known to be stable close to the anti-continuum limit for the
focusing nonlinearities considered herein~\cite{book}. 
In the present case, its corresponding eigenvalues are
found to be
$\lambda_{1,2}=\pm \sqrt{8 \epsilon} i$, 
$\lambda_{3,4,5,6}=\pm 2 \sqrt{\epsilon (1-k)} i$ (a double pair),
while finally there is a pair at the origin, as expected due to the 
relevant invariance. Notice that in this case an instability
is predicted when the NNN interaction strength overcomes the NN
one i.e., for $k>1$. We will return to this effect later in our exposition.

For now, let us comment on the very good agreement of the above
predictions with what is shown in the left panel of Fig.~\ref{lfig1a}
for $\epsilon=0.005$ (that is used hereafter).
Furthermore, the case of $k=1.5$ is selected on the right panel to
indicate that although the solution is stable in the NN limit, a
sufficiently strong NNN interaction may destablize it, leading
to an amplitude symmetry breaking exchange of power among the 4 principal
sites.  

\begin{figure}[tbp]
\begin{center}
\epsfig{figure=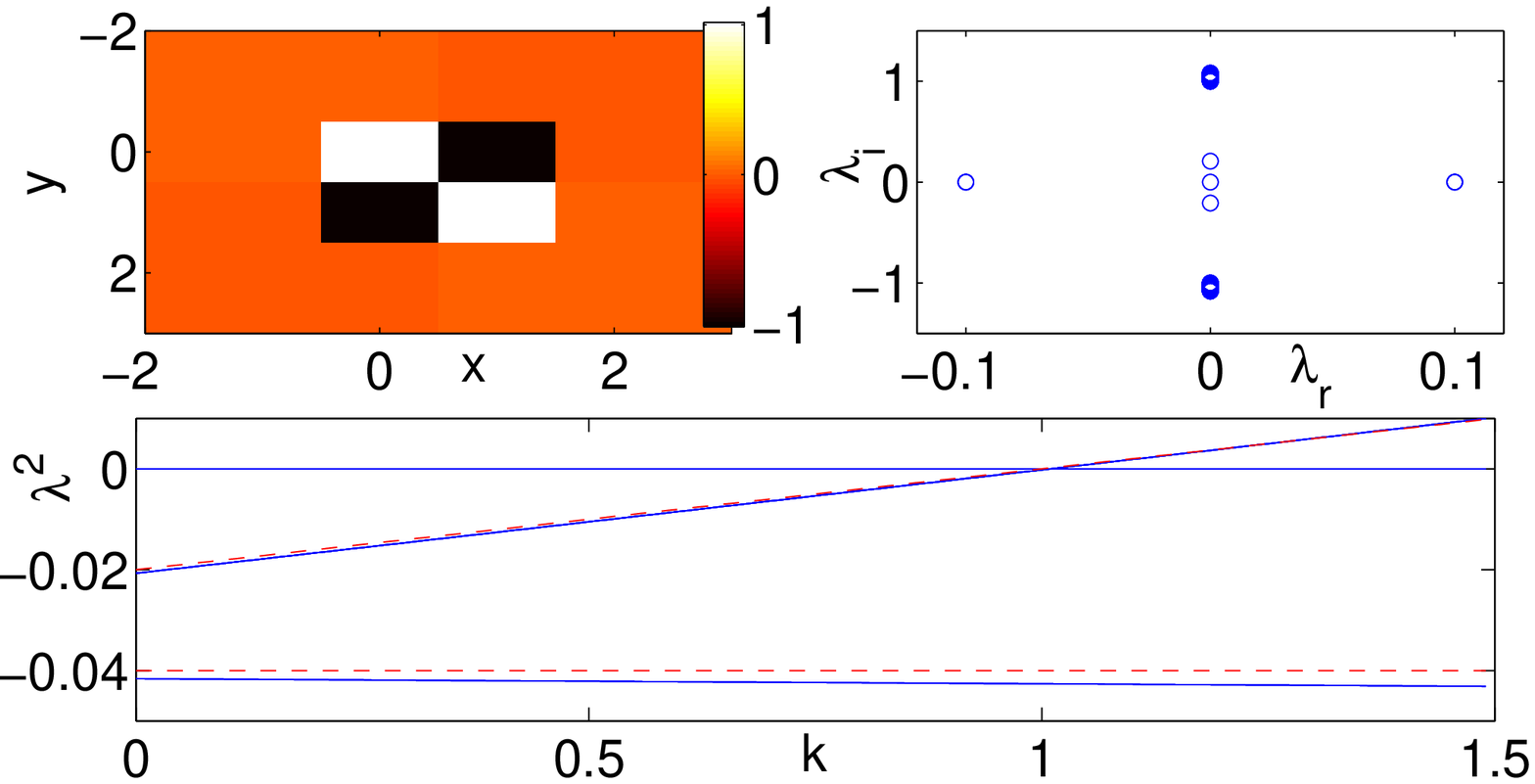,height=2in,width=3in,angle=0}
\epsfig{figure=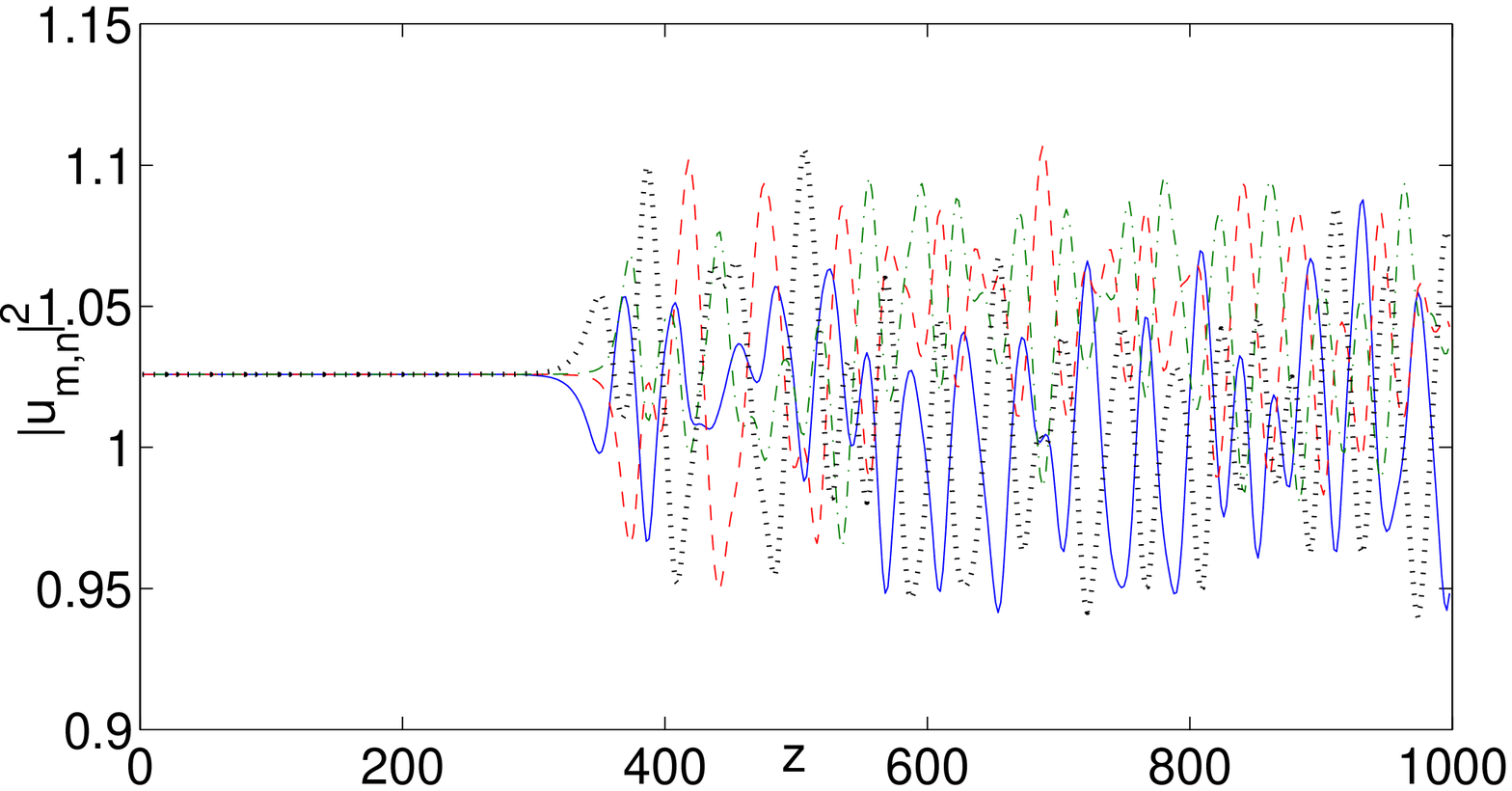,height=2in,width=3in,angle=0}
\end{center}
%\begin{center}
%\epsfig{figure=LRI_new_NN4.eps,height=2in,width=5in,angle=0}
%\epsfig{figure=LRI_new_NN3.eps,height=2in,width=5in,angle=0}
%\end{center}
\caption{The left top panels correspond to the case of the
$(0,\pi,0,\pi)$ state with $k=1.5$. The instability of this
branch for this supercritical case is evident in the spectral
plane of the linearization. The bottom left panel corresponds
to the continuation over the NNN interaction strength $k$
and the squared eigenvalue ($\lambda^2$) zero-crossing corresponds to the
destabilization of the branch. Notice that we will often use hereafter
this diagnostic ($\lambda^2$) in cases devoid of complex eigenvalues,
as its zero crossings are characteristic of stability changes of the
solution and indicative of potential bifurcation points.
This destabilization is
dynamically illustrated in the right panel for $k=1.5$,
again featuring for sufficiently long propagation distances
an amplitude symmetry breaking and an intensity oscillation
of the 4 principal sites.}
\label{lfig1a}
\end{figure}

\item Another principal, yet highly unstable configuration of the
square contour that is predicted from our solvability conditions to
persist is that with ${\bf \theta}=(0, 0, 0, 0)$. Here all the $\phi$'s
are $0$. We examine this case mostly for completeness (and also because
of its potential relevance and stability 
for the defocusing case of $\epsilon<0$).
The corresponding eigenvalues here are given by 
$\lambda_{1,2}=\pm \sqrt{8 \epsilon} $, 
$\lambda_{3,4,5,6}=\pm 2 \sqrt{\epsilon (1 + k)} $ (a double pair),
as well as a pair of zero eigenvalues. It should be noted here
that this result, once again found to be in excellent agreement
with our numerical computations in the left panels of Fig.~\ref{lfig2},
suggests a partial restabilization of this branch when the double
pair becomes imaginary for $k<-1$. On the other hand, the dynamics
of the branch shown in the right panel of the figure, is interesting
in its own right as it suggests a pairing of $(0,0)$ and $(1,0)$
in an oscillatory pattern and of $(0,1)$ and $(1,1)$ in a similar pattern
[although this appears to change for sufficiently long time scales].

\begin{figure}[tbp]
\begin{center}
\epsfig{figure=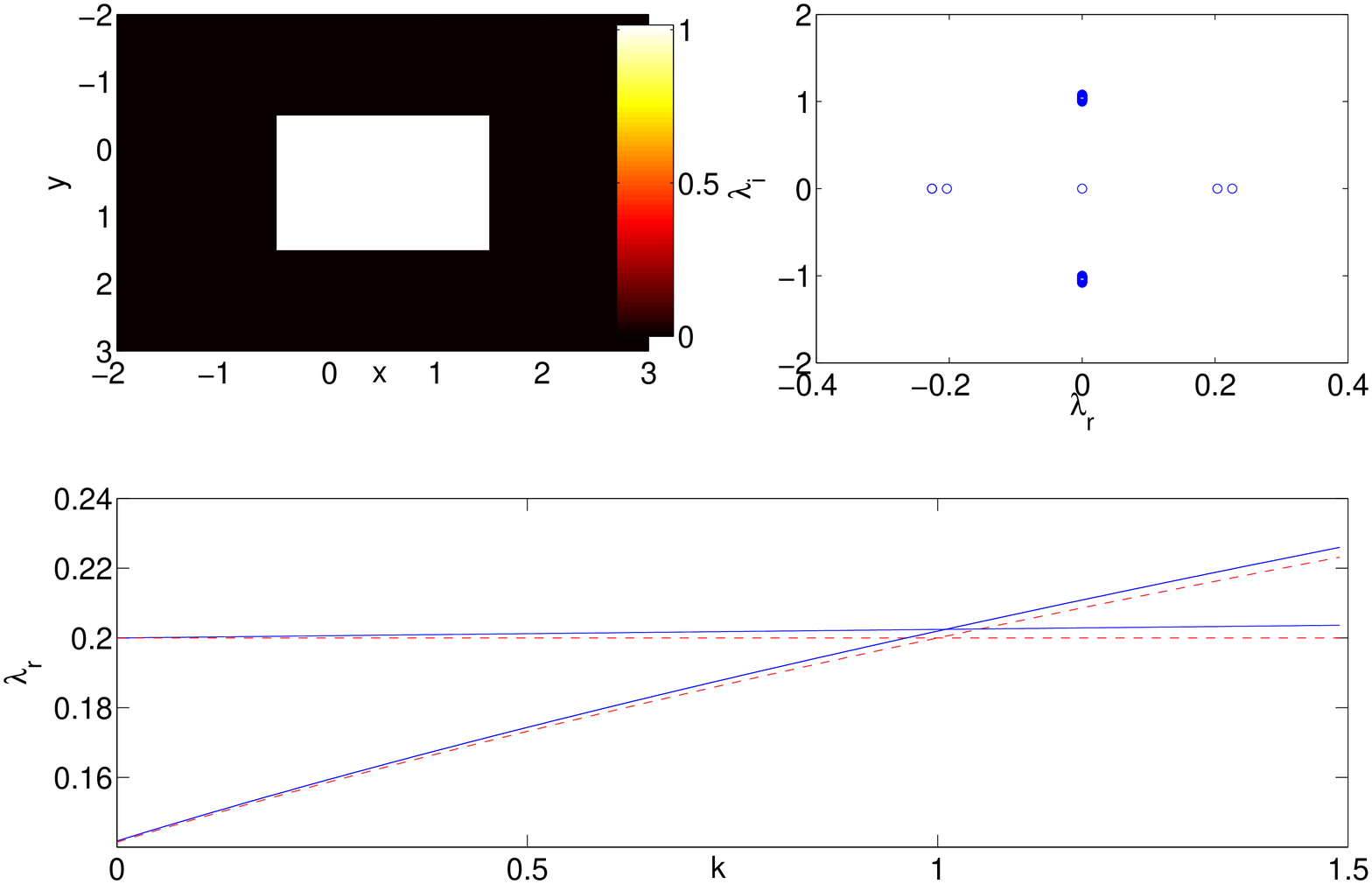,height=2in,width=3in,angle=0}
\epsfig{figure=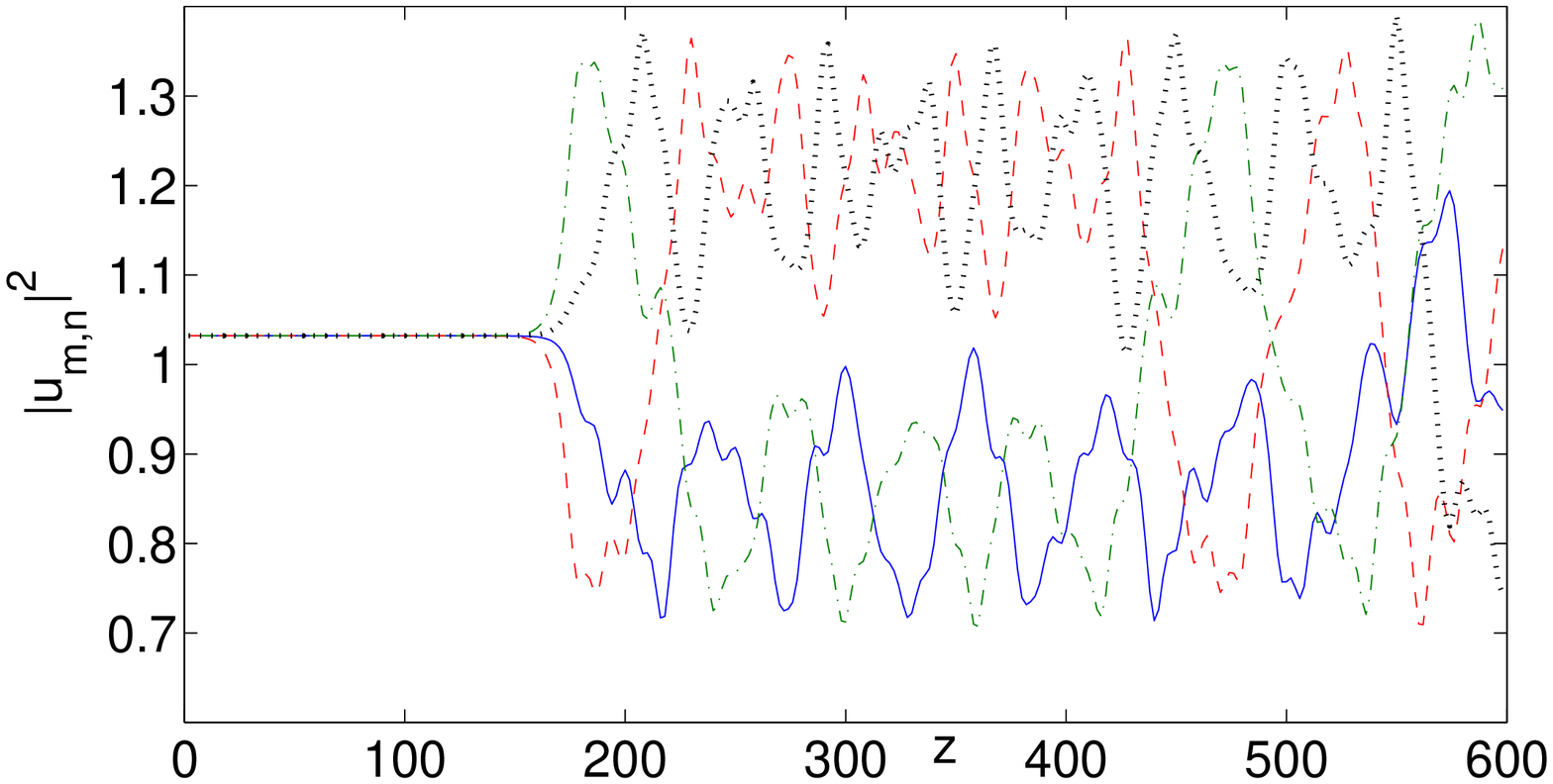,height=2in,width=3in,angle=0}
\end{center}
\caption{The left panel of the figure shows the continuation and
stability analysis of the ${\bf \theta}=(0,0,0,0)$ branch. The top
panels show an example of this branch for $k=1.5$, while the bottom
panel shows the corresponding single and double eigenvalue pairs.
The right panel shows the evolution of the four central sites
of the configuration (in the same way as before) for $k=1.5$, 
indicating their
paired oscillations between  $(0,0)$ and $(1,0)$ and separately
$(0,1)$ and $(1,1)$ for a lengthy interval during the propagation.}
\label{lfig2}
\end{figure}

\item The next example is that of ${\bf \theta}=(0, 0, \pi, 0)$,
for which $\phi_1=0$, while $\phi_2=\phi_3=\pi$. In this case,
our analytical calculation of the eigenvalues yields (in addition
to the null pair) a pair at $\lambda_{1,2}=\pm 2 \sqrt{k \epsilon}$,
a separate one at $\lambda_{3,4}=\pm \sqrt{2 \epsilon}
\sqrt{k + \sqrt{8 + k^2}} i$ and one at $\lambda_{5,6}=\pm \sqrt{2 \epsilon}
\sqrt{-k + \sqrt{8 + k^2}}$. A brief inspection of these eigenvalue
pairs confirms that there should always be at least 1 real and positive
eigenvalue associated with this solution (for the focusing case,
under study). If $k<0$, then there is exactly one such eigenvalue, while
if $k>0$, there are two real eigenvalues.
Hence, it should always be unstable.
This, as well as our detailed prediction for the dependence of the $\lambda$'s
on $k$, are very accurately reflected in the full numerical computations of
Fig.~\ref{lfig3}. The left panel shows a prototypical example of the state
and its systematic continuation over $k$, while the right panel illustrates
the oscillatory (yet not clearly periodic) pattern of exchange of 
intensity, upon the amplitude 
symmetry breaking that signals the pattern's predicted
instability for $k=1.5$.

\begin{figure}[tbp]
\begin{center}
\epsfig{figure=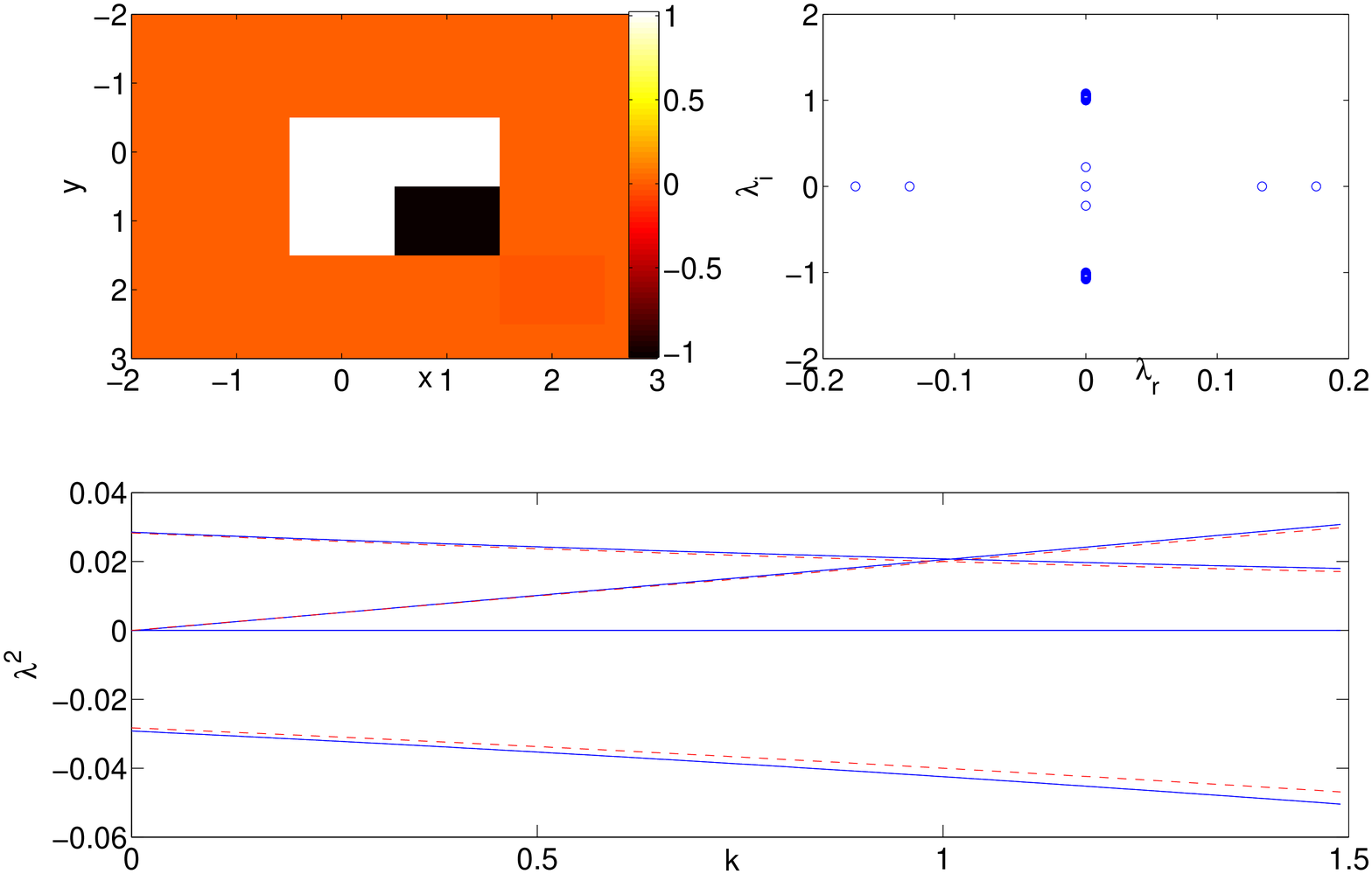,height=2in,width=3in,angle=0}
\epsfig{figure=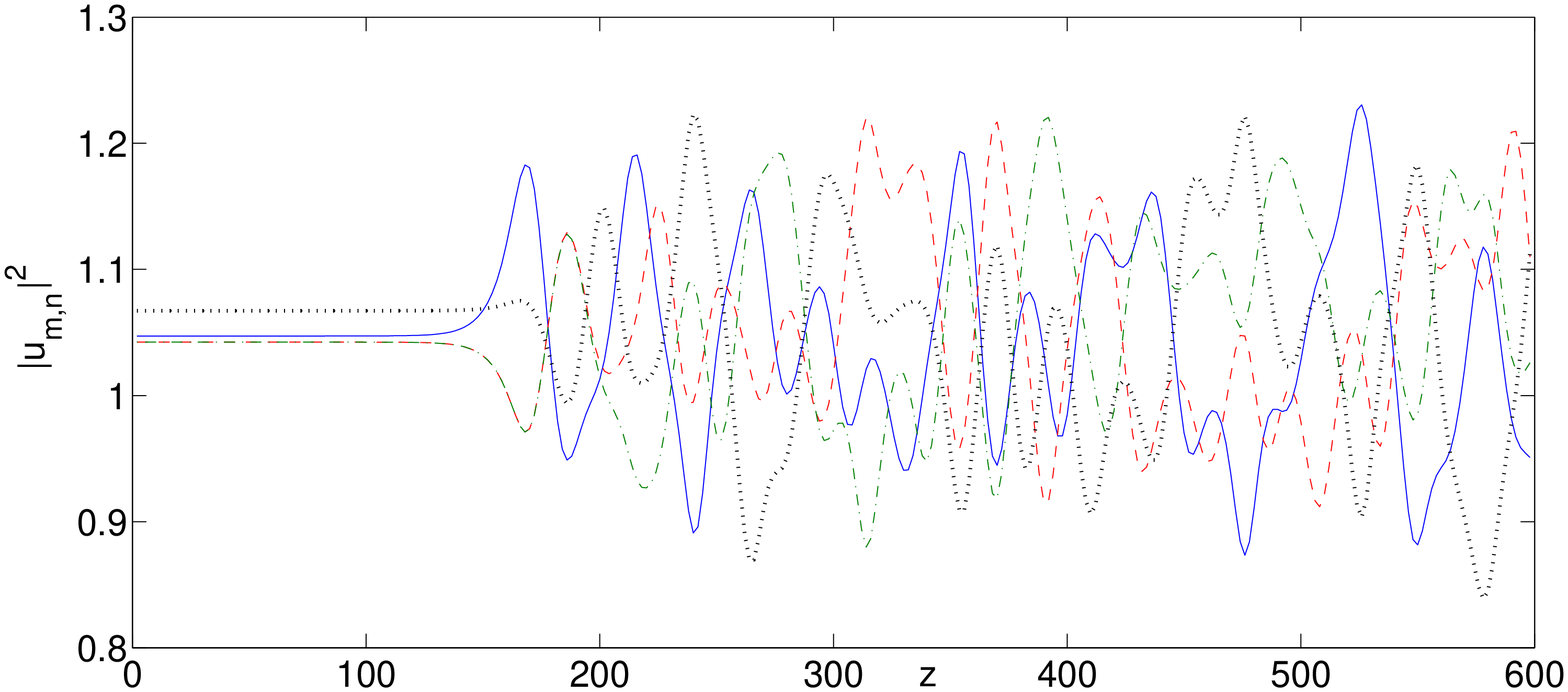,height=2in,width=3in,angle=0}
\end{center}
\caption{The left panel shows a case example of the 
${\bf \theta}=(0, 0, \pi, 0)$ configuration and its corresponding
spectral plane (top). In the bottom, the dependence of the associated
eigenvalues (see text) on $k$ is given. The right panel shows the
evolution of the central site intensities for the unstable 
configuration at $k=1.5$.}
\label{lfig3}
\end{figure}

\item We now turn to the case of ${\bf \theta}=(0, \pi, \pi, 0)$,
for which $\phi_1=\phi_3=\pi$, while $\phi_2=0$. Once again,
the eigenvalues of the linearization can be computed to leading
order yielding in this case a double pair at the origin,
one pair which is real in the absence of beyond-nearest-neighbor
interactions and becomes modified according to
$\lambda_{5,6}=\pm 2 \sqrt{ \epsilon (1-k)}$, in the presence
of the $k$-dependent next-nearest-neighbor interaction.
Notably, this dependence leads to restabilization of the 
configuration for $k>1$. Finally, the 4th pair is imaginary
for $k>0$ (but can lead to -further- destabilization for $k<0$), according
to $\lambda_{7,8}=\pm 2 \sqrt{\epsilon (1+k)} i$. 
Fig.~\ref{lfig4} confirms once again the excellent agreement of the
theoretical predictions with the relevant eigenvalue results
(see the top right panel) and the existence of the instability
in the absence of or for sufficiently weak beyond nearest neighbor
interactions; see top left and bottom left panels. On the other
hand, it also confirms the dynamical stability for the case of $k=1.5$
in the top left and bottom right panels. In the latter the small
perturbation leads to bounded oscillatory dynamics, instead of the
unstable evolution of $k=0$ (bottom left).

\begin{figure}[tbp]
\begin{center}
\epsfig{figure=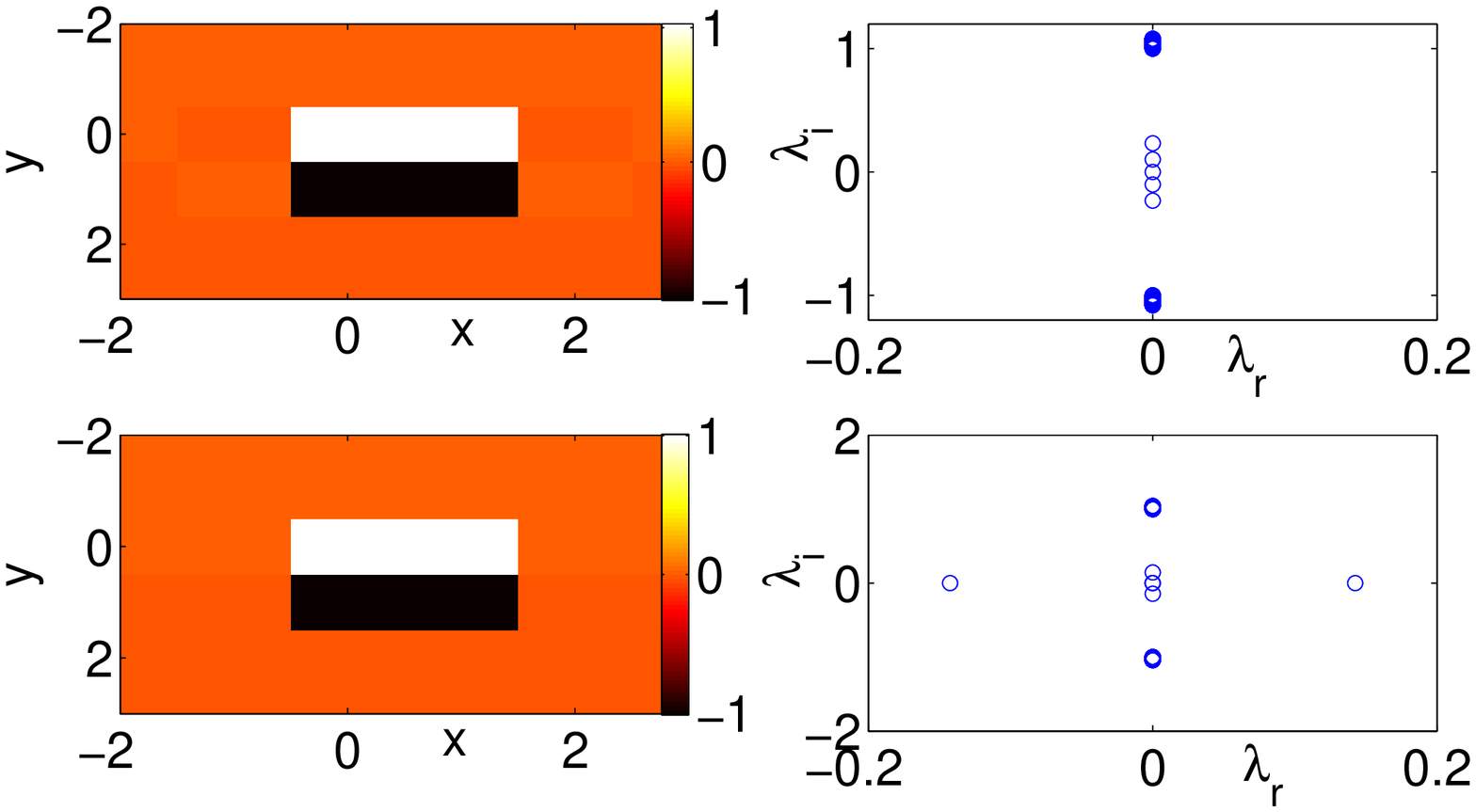,height=2in,width=3in,angle=0}
\epsfig{figure=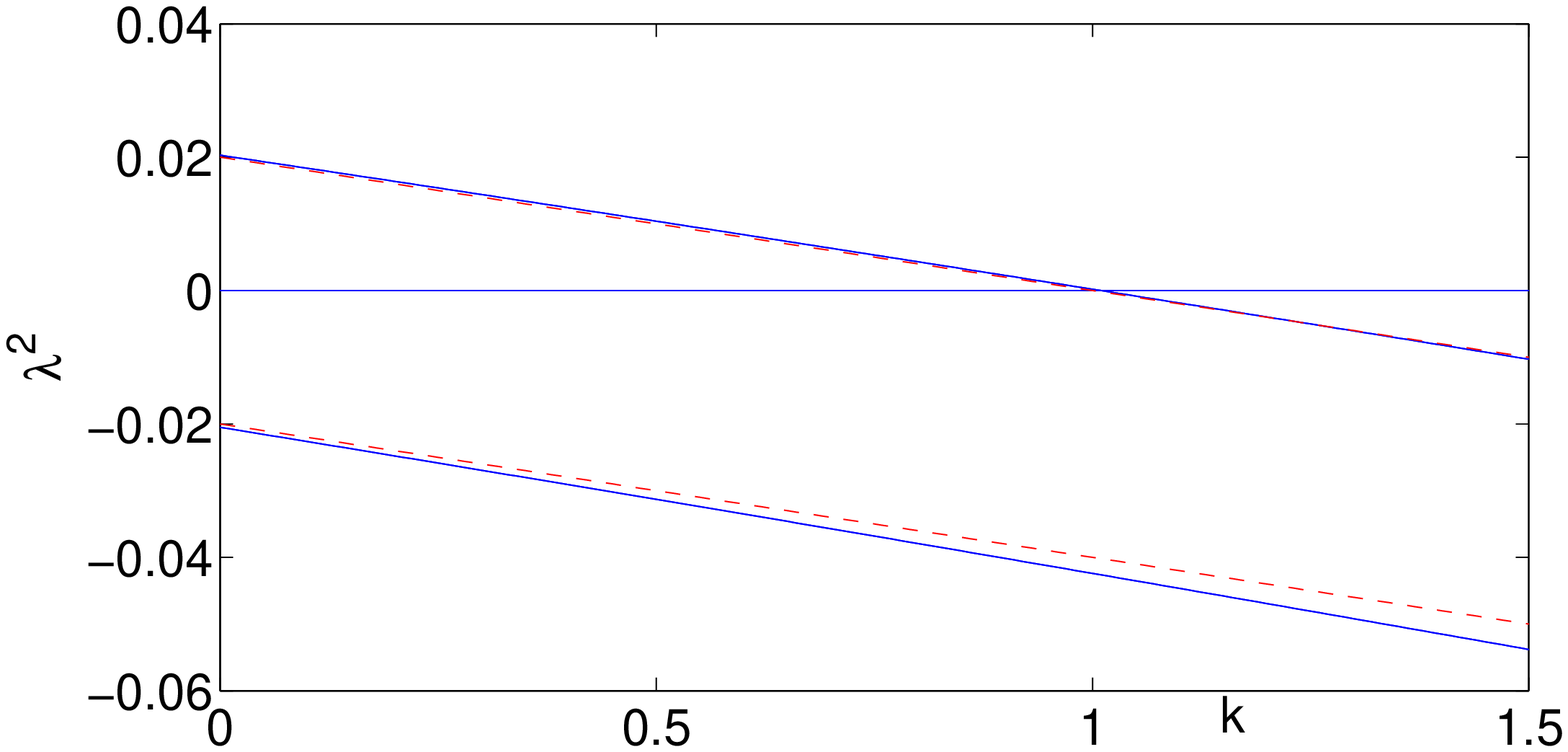,height=2in,width=3in,angle=0}
\end{center}
\begin{center}
\epsfig{figure=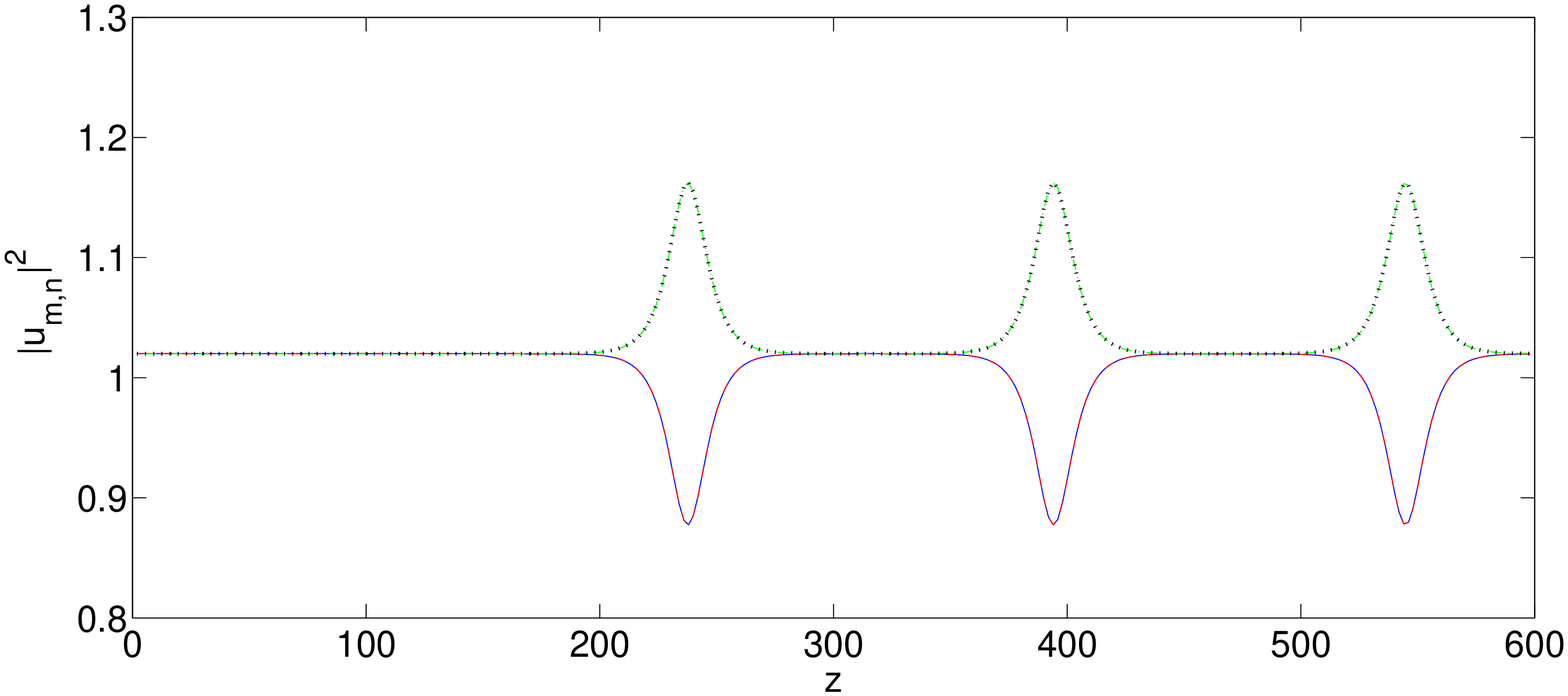,height=2in,width=3in,angle=0}
\epsfig{figure=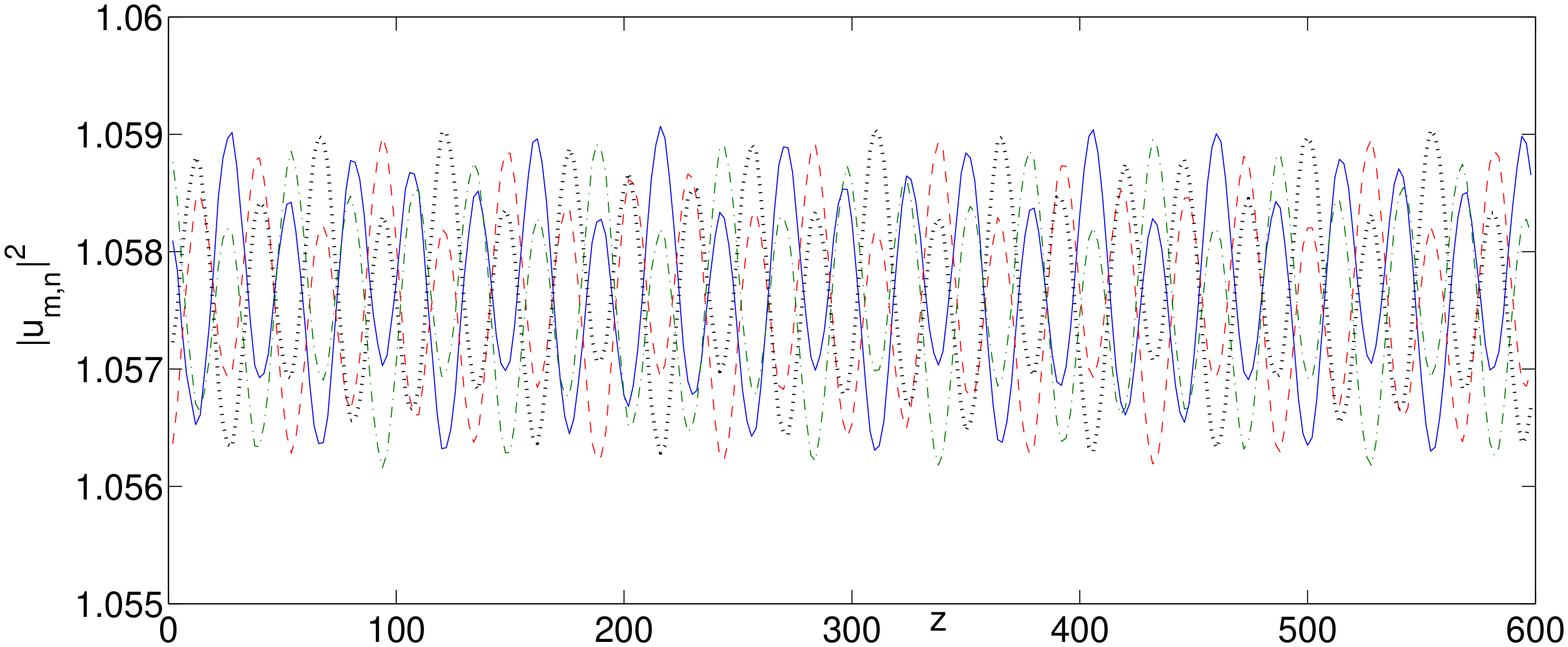,height=2in,width=3in,angle=0}
\end{center}
\caption{The top left panel shows a case example of the 
${\bf \theta}=(0, \pi, \pi, 0)$ for the stable case of $k=1.5$
and the unstable one of $k=0$. The top right panel 
confirms this transition
from instability to stability as $k$ is increased through following
the corresponding eigenvalues numerically (blue solid) and analytically
(red dashed lines), in excellent agreement betwen the two. The dynamical
evolution of the principal 4 sites is demonstrated in the bottom 
left panel for
the unstable $k=0$ scenario, giving rise to a periodic emergence of an
asymmetric pattern in the intensity of the sites. Finally, the bottom
right panel case of $k=1.5$ only leads to (small fluctuation amplitude)
bounded oscillatory dynamics even
when perturbed, confirming its predicted dynamical stability.}
\label{lfig4}
\end{figure}

\item It is especially interesting to note that the 
configurations ${\bf \theta}=(0, \pi, \pi, 0)$
and ${\bf \theta}=(0, \pi, 0, \pi)$ in the limit of
$k=1$ become equivalent to each other. This is a byproduct
of the equal strength of interaction of each of the sites
with any one of its 3 (nearest or next nearest) excited
neighbors. In this special limit, for both of these
configurations, each of the excited phases of $0$ ``sees''
a neighbor with the same phase and two neighbors with $\pi$
phase and each of the $\pi$ phase excited sites ``sees'' another
$\pi$ and two $0$ phases, rendering the configurations equivalent.
This is manifested also by the equality of their respective eigenvalues
in the expressions given above (they share a triple pair of 0's
and one pair of $\sqrt{8 \epsilon} i$). Given the stability change
of these configurations at $k=1$ [${\bf \theta}=(0, \pi, \pi, 0)$ 
transitions from instability to stability as $k$ increases through the
unit value, while ${\bf \theta}=(0, \pi, 0, \pi)$ transitions in
the opposite direction], we expect a potential bifurcation of
a new branch past this critical point. Indeed, this is precisely
what the analytical formulas of Eqs.~(\ref{peq1})-(\ref{peq3}) 
predict. More specifically, the longer range interactions
considered herein are not only responsible for stability changes
(or exchanges), but additionally lead to the formation of entirely
new branches of solutions that would be {\it absent in the nearest
neighbor limit}. Such a branch is given by 
${\bf \theta}=(0, \cos^{-1}(-1/k), 2 \cos^{-1}(-1/k), \cos^{-1}(-1/k))$.
In this case, $\phi_1=\phi_2=-\phi_3=\cos^{-1}(-1/k)$.

We can use the Jacobian formulation to provide explicit analytical
predictions of the corresponding eigenvalues in this case, as well.
In particular, in addition to the standard pair of eigenvalues at the origin,
the other 3 pairs are non-vanishing; $\lambda_{3,4}=\pm \sqrt{8 \epsilon/k} i$,
while $\lambda_{5,6}=\pm 2 \sqrt{\epsilon (1-k^2)/k} i$
and  $\lambda_{7,8}=\pm 2 \sqrt{\epsilon (1-k^2)/k}$. From the above,
it is clear that among the two extra vanishing eigenvalue pairs
(at $k=1$) of the configurations  ${\bf \theta}=(0, \pi, \pi, 0)$
and ${\bf \theta}=(0, \pi, 0, \pi)$, in this ``double pitchfork''
bifurcation, one always exits as real and one exits as imaginary.
This is a degenerate pitchfork bifurcation because at the critical
point, there exist two eigenvalue pairs at the $\lambda=0$, which
for $k>1$ move in different directions. As a result, this novel
configuration created solely by the beyond-nearest-neighbor interactions
will generically be found to be unstable in its interval of existence.
These theoretical predictions are fully confirmed
in Fig.~\ref{lfig5}. In particular, very good agreement (with the
above theory) is obtained for
the 3 pairs of bifurcating eigenvalues in the left panel, and the instability
of the configuration with $k=3$ (left panel) is confirmed in the direct
numerical simulations of the right panel.

\begin{figure}[tbp]
\begin{center}
\epsfig{figure=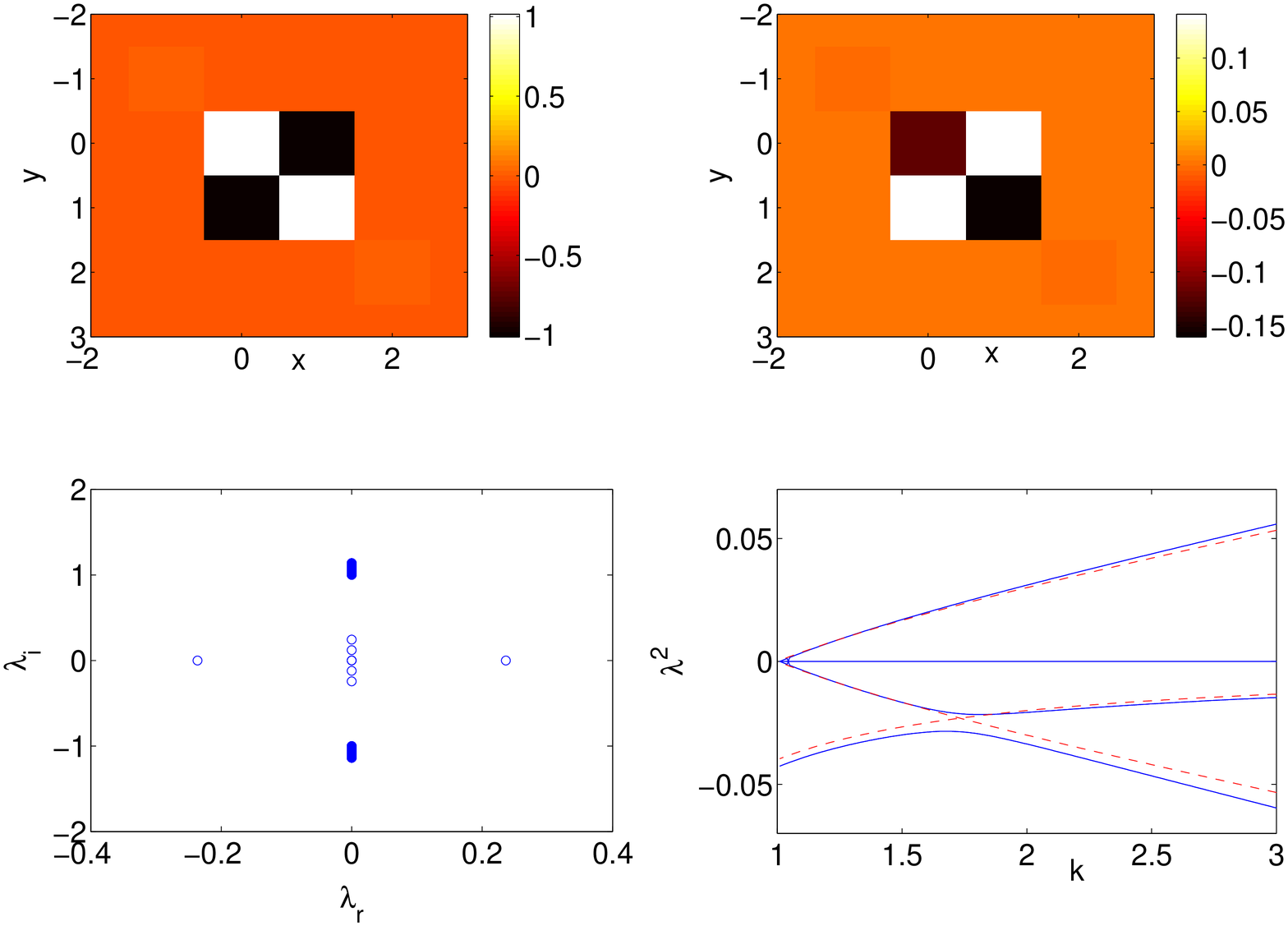,height=2in,width=3in,angle=0}
\epsfig{figure=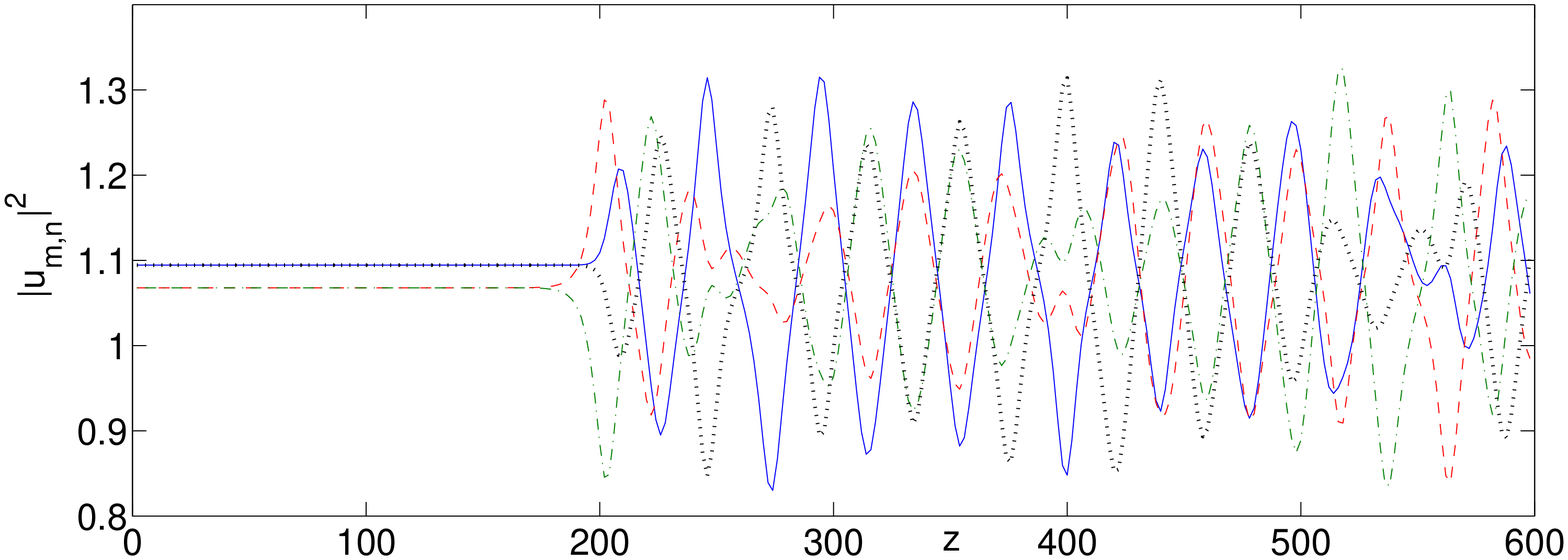,height=2in,width=3in,angle=0}
\end{center}
\caption{The left panel shows the (unstable) configuration with phase
distribution ${\bf \theta}=(0, \cos^{-1}(-1/k), 2 \cos^{-1}(-1/k), \cos^{-1}(-1/k))$ for $k=3$. It also shows the dependence of the numerical eigenvalues
(blue solid lines) for this branch and their comparison to the theoretical
predictions (red dashed lines). The right panel confirms the instability
of the configuration of the left panel for $k=3$, 
given the observed (strong)
amplitude symmetry breaking
in the intensities of the four sites from their (unstable) equilibrium
values.}
\label{lfig5}
\end{figure}

\item Finally, we touch upon a branch of solutions that is predicted
by the leading order expansion and can be obtained for the values
of $\epsilon$ used here (in particular, for $\epsilon=0.005$), but
which does not exist as an exact solution in the case of $k=0$
and hence we believe does not exist here either. Nevertheless, we have
not proved this rigorously, since the proof would necessitate resorting
to sufficiently high order expansions. We only infer this from the
need to lower our tolerance to obtain the relevant solution (and the
rigorous proof of its non-existence in~\cite{peli_2d} for the $k=0$ case,
despite its theoretical proposition in physical setups in~\cite{sukh}).
Such asymmetric vortices have the phase profile 
${\bf \theta}=(0, a, \pi, \pi+a)$, with $\phi_1=a$, $\phi_2=\pi-a$
and $\phi_3=a$. At the level of the leading order reductions, it
is predicted that such vortices have a double pair of zero eigenvalues,
while the other two pairs are located at
$\lambda_{5,6}=\pm 2 \sqrt{\epsilon} \sqrt{\cos(a)-k}$
and $\lambda_{7,8}=\pm 2 \sqrt{\epsilon} \sqrt{\cos(a) + k} i$.
It is noteworthy that if this branch was an exact solution, it would
naturally generalize the vortex branch with $a=\pi/2$
and the mixed phase branch ${\bf \theta}=(0, \pi, \pi, 0)$
for $a=\pi$. It is then also natural to expect the extra
vanishing pair of eigenvalues of such branches, due to essentially
the invariance of the branch with respect to $a$. Nevertheless,
as was shown e.g. for $a=\pi/2$ and $k=0$, this extra pair does
not stay put at the origin, but instead it bifurcates at a higher order.
The relevant ``approximate'' branch of solutions is shown in 
Fig.~\ref{lfig6} for $a=\pi/8$. We can see that it is always unstable
for the range of considered $k$'s. This instability does
appear to lead to symmetry-breaking pairwise oscillations/mass exchanges 
between the excited sites in the right panel of the figure. 
Although it is predicted that
the instability should disappear for high enough $k$'s (beyond
$k=\cos(a)$), we were unable to converge to the solution up to
that value of $k$ (for $a=\pi/8$), even
for the reduced tolerance of $10^{-5}$ for the error 
in the convergence to the solution of our fixed point iteration used 
for this branch. While this is not conclusive in any way, 
it may be suggestive since these problems did not arise for 
the cases of $a=\pi/2$ or $a=\pi$ examined above.

\begin{figure}[tbp]
\begin{center}
\epsfig{figure=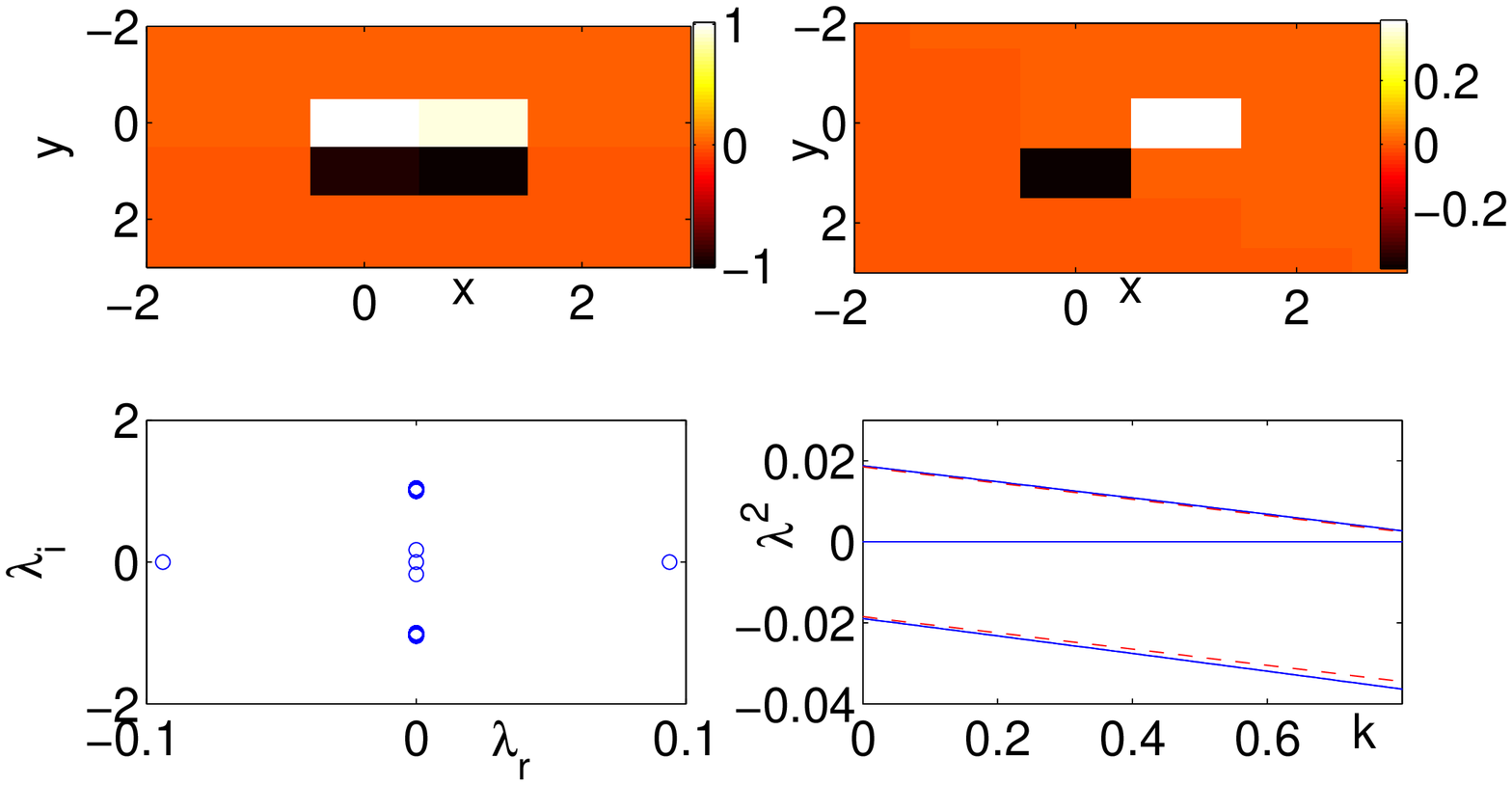,height=2in,width=3in,angle=0}
\epsfig{figure=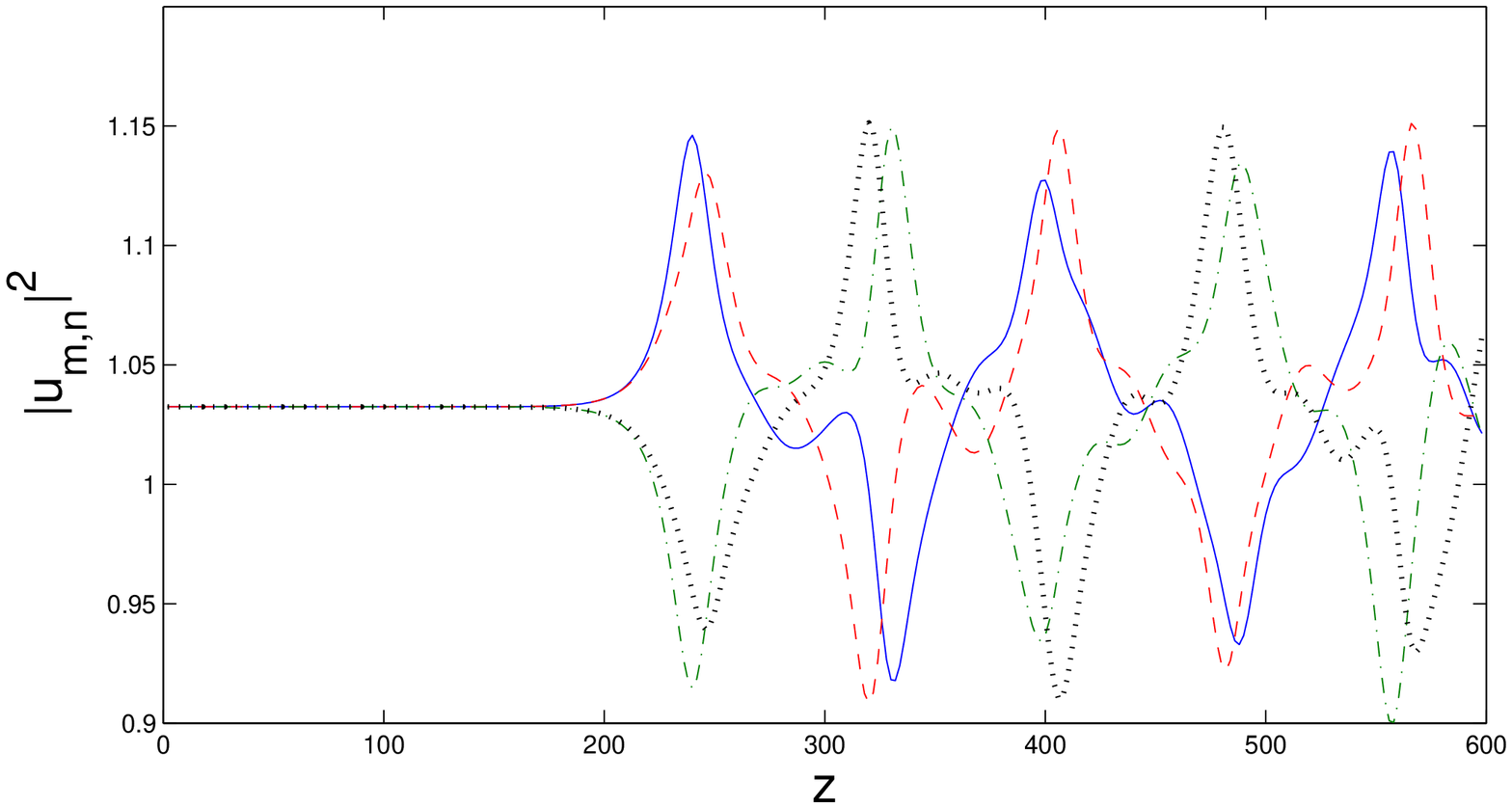,height=2in,width=3in,angle=0}
\end{center}
\begin{center}
\end{center}
\caption{A case example of the approximate asymmetric solution
branch with $a=\pi/8$, shown in the left panel for $k=0.5$. Its
eigenvalues appear to be in very good agreement with the theoretically
predicted ones (as a function of $k$), yet the solution is only
approximate given that we have had to reduce the fixed point iteration
tolerance to ``converge'' to it. The right panel shows the
dynamical evolution of the solution at the left showcasing its 
lack of dynamical robustness due to a nearly periodic exchange of
power between each of the two pairs of excited sites.}
\label{lfig6}
\end{figure}

\end{enumerate}

\begin{figure}[tbp]
\begin{center}
\epsfig{figure=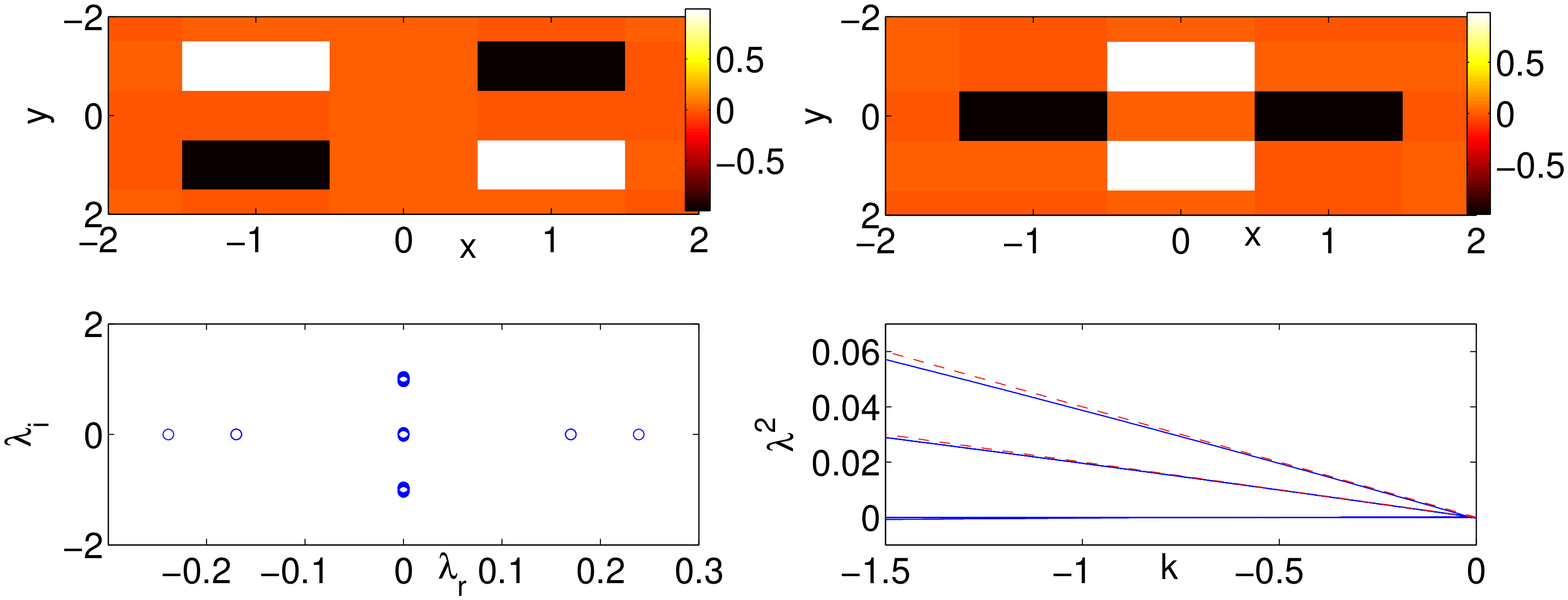,height=2in,width=3in,angle=0}
\epsfig{figure=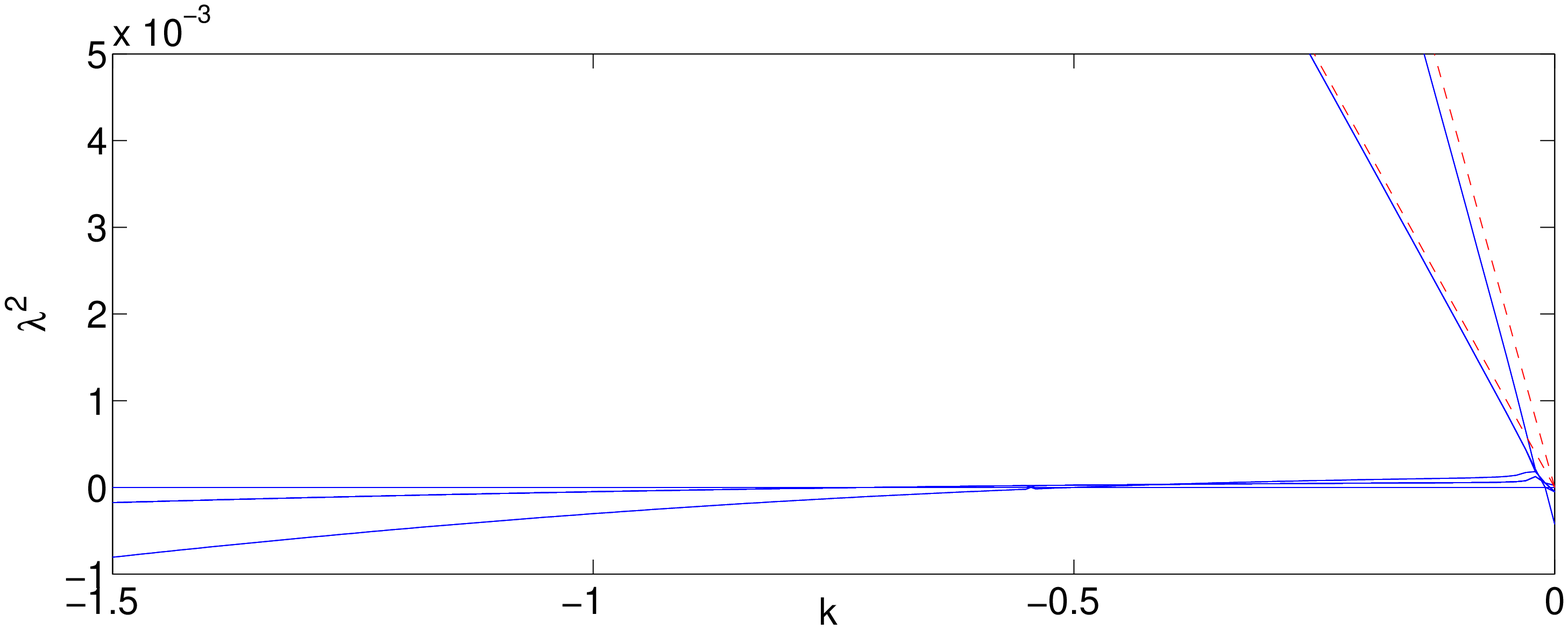,height=2in,width=3in,angle=0}
\end{center}
\begin{center}
\epsfig{figure=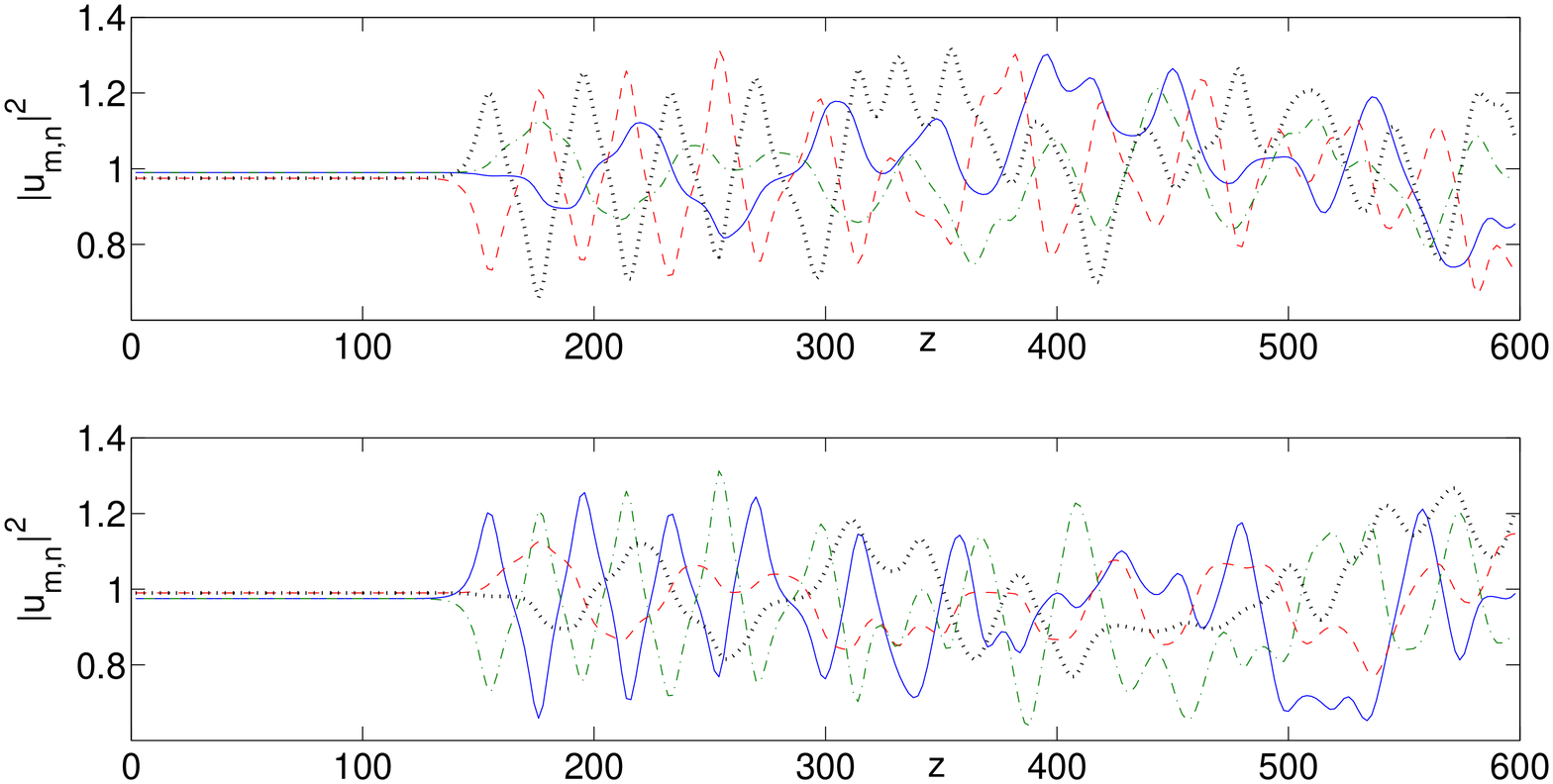,height=2in,width=3in,angle=0}
\epsfig{figure=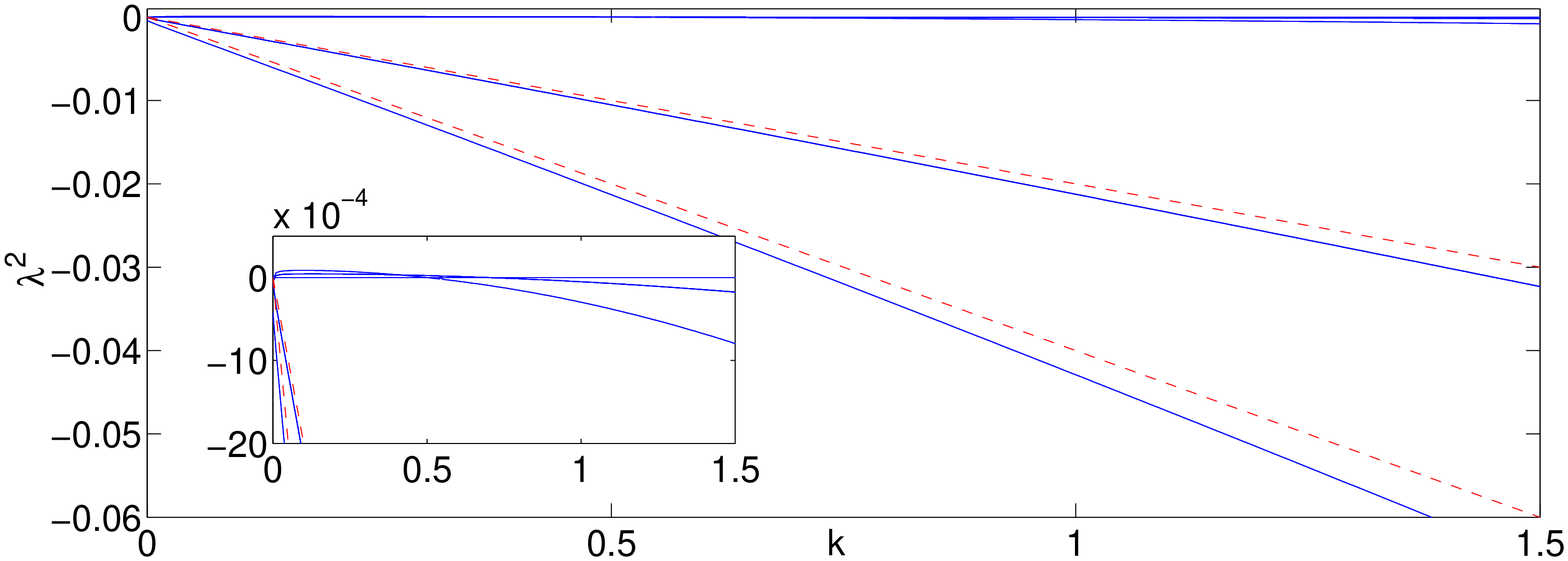,height=2in,width=3in,angle=0}
\end{center}
\caption{The top left panel presents the case example of the vortex of
topological charge $S=2$ for the unstable case of $k=-1.5$. The relevant
dominant eigenvalues shown also can be discerned in the blow-up of the
top right panel that showcases more clearly the lower order eigenvalues which
become stabilized due to the effect of $k$. Nevertheless, as shown in the
evolution of the bottom left panel, the instability for $k=-1.5$ cannot
be avoided and is manifested in the asymmetric evolution of the 8 sites
participating in the vortex (split into two pairs of four sites,
namely the lower left ones and the top right ones). On the other hand, the bottom right panel
shows that $k>0$ has the same beneficial effect for the higher order
eigenvalues (they become stabilized, and all have crossed $\lambda^2=0$
by the value of $k \approx 0.7$). Yet in this case, also the leading order
eigenvalues (well captured by our theory of the dashed red lines) are
now imaginary and hence the vortex of $S=2$ is completely dynamically 
stabilized.}
\label{lfig7}
\end{figure}

Finally, to illustrate the powerful nature of the beyond nearest
neighbor interactions as a ``controller'' of not only the existence
but also the stability of complex nonlinear wave configurations,
in Fig.~\ref{lfig7}, we present a select example from a configuration
of a larger (8-site) contour in the case of a vortex of
topological charge $S=2$. In the case of $k=0$, the stability of
this vortex has been analyzed in~\cite{peli_2d} (see also the earlier
numerical investigations of~\cite{pgk4}) and it was found
that it was always unstable due to a higher order eigenvalue
(proportional to $\epsilon$). In the present case, including the
beyond nearest neighbor terms, leads to a dominant order prediction
of 5 eigenvalue pairs  at the origin, a double pair given
by $\lambda_{1,2,3,4}=\pm 2 \sqrt{\epsilon k} i$ and a single
pair of $\lambda_{5,6}=\pm \sqrt{8 \epsilon k} i$. In Fig.~\ref{lfig7},
we observe that for $k<0$, these dominant eigenvalues give rise
to a strong destabilization (with $\lambda^2 \propto \epsilon$,
i.e., stronger than the $k=0$ case) of the coherent structure.
The associated instability is also evidenced dynamically in the
figure for $k=-1.5$.
On the other hand, for $k>0$, the relevant eigenvalues at O$(\sqrt{\epsilon})$
are imaginary and we observe from the figure that also all the higher
order eigenvalues cross the stabilization threshold of $\lambda^2=0$
and become imaginary (the last pairs cross for $k \approx 0.7$). As
a result, the increase of the next-nearest-neighbor interaction
is responsible for the {\it complete stabilization} of the
vortex of topological charge $S=2$. In that light, we conclude,
that not only are such beyond-nearest-neighbor interactions potentially
responsible for the destabilization of states that were stable
in the nearest-neighbor-interaction limit (such as the vortex of $S=1$).
They are also potentially responsible for the stabilization of
unstable states of that limit such as the vortex of $S=2$. Finally,
as we illustrated above, they are also responsible for the
emergence of novel states (such as the branch in item 6 above)
and of unusual bifurcations (such as the double pitchfork that
we obtained at the limit of $k=1$).

\section{Conclusions and Future Challenges}

In this paper, we have focused our interest on discrete
two-dimensional dynamical lattices of the DNLS type, which are
a prototypical model for a variety of potential applications,
including waveguide arrays in nonlinear optics and BECs
in optical lattices (in the superfluid regime) in atomic
physics. The principal theme of the study was the role
of beyond-nearest-neighbor interactions on the prototypical
results that are known and understood for the more standard
and more extensively studied case of the nearest-neighbor
interaction~\cite{book,peli_2d}. 

What was found was that such additional interactions may play
a critical role in shaping the associated dynamics. This is
evident by their ability to destabilize stable configurations
(such as the 4-site vortex of topological charge $S=1$)
and also their potential to stabilize unstable configurations
(such as the 8-site vortex of topological charge $S=2$), for
sufficient strength and suitable signs of these beyond-nearest-neighbor
effects. These two above mentioned examples are perhaps 
particularly notable because they belong to the category
of the so-called super-symmetric states of~\cite{peli_2d}.
For such states (for which the relative phase between 
adjacent excited sites in the contour is $\pi/2$), the contribution 
to the linearization Jacobian of the solvability conditions that
we used to compute the full problem eigenvalues, remarkably, vanishes.
Hence, in the nearest neighbor limit the dominant eigenvalues of such
super-symmetric states arise with $\lambda$ of O$(\epsilon)$ or
higher. In that light, the inclusion of beyond nearest neighbor
interactions yields an effect which is {\it dominant} to leading order
(with $\lambda^2 \propto k \epsilon$) even when $k$ is 
small. It is thus natural to expect that 
especially in such super-symmetric cases, these longer range
interactions play the role of a powerful controller affecting
the potential stability of the ensuing nonlinear wave states.

In addition to that possibility, we obtained a series of
``solitonic'' solutions (without a vortex structure), which
in their own right had some interesting stability modifications,
for sufficiently strong beyond nearest neighbor interactions.
There, too, we saw solutions like the $(0,\pi,0,\pi)$ start
out as stable for small $k$ but become unstable for large
values of $k$, and vice versa solutions like $(0,\pi,\pi,0)$
which start as unstable but are stabilized as $k$ increases.
In fact, these two solutions participate in an intriguing
double pitchfork bifurcation at the degenerate limit 
of equal nearest and next nearest neighbor interactions of
$k=1$. Off of this limit, we observe the potential of
the next-nearest-neighbor terms to produce novel states
such as those produced in item 6 above.
Hence, this powerful controller of the beyond nearest neighbor
interactions is responsible for the emergence of previously
unfeasible wave states.

We believe that through this prototypical example, we have
made the case for the substantial relevance and interest
within the consideration of interactions that go beyond the
nearest neighbor effects. This is especially so in particular configurations
(such as the super-symmetric ones) where the nearest neighbor effects
are not discernible and hence higher order interactions are dominant
even as soon as they arise. It is thus an interesting direction
to try to appreciate their effects more systematically, 
for different kernels, such as Gaussian, or exponentially 
decaying ones~\cite{review_lri}. On the other hand, it would
seem especially interesting to try to generalize relevant consideration
to higher dimensions and to 3-dimensional solitons, vortices
and vortex cubes~\cite{peli_3d,book}, to obtain a systematic
view of beyond nearest neighbor interactions in such settings, as well.
Such studies are deferred to future publications.

%\begin{figure}
%\begin{center}
%\begin{tabular}{cc}
%    (a) & (b) \\
%\epsfxsize=8.25cm %\centerline{}
%\epsffile{dwb4.ps} &
%\epsfxsize=8.25cm %\centerline{}
%\epsffile{dwb4a.ps} \\
%    (c) & (d) \\
%\epsfxsize=8.25cm %\centerline{
%\epsffile{dwb5.ps} &
%\epsfxsize=8.25cm %\centerline{}
%\epsffile{dwb7c.ps} \\
%\end{tabular}
%\caption{Panel (a) shows the profile and spectral plane of a
%potential $V(x)=0.025 x^2 + 2 \cos(2 x)$.}%
%\label{dfig2}
%\end{center}
%\end{figure}

\vspace{5mm}

{\it Acknowledgements}: The author is grateful to the US National
Science Foundation for support under grants NSF-DMS-0806762,
NSF-CMMI-1000337 and to the US-AFOSR for support under grant
FA9550-12-1-0332, as well as to the Alexander von Huboldt,
Alexander S. Onassis (grant RZG 003/2010-2011) and Binational
Science (grant 2010239) Foundations. He also gratefully acknowledges
Dr. V. Koukouloyannis for
numerous discussions on the theme of longer range interactions.

%\end{multicols}

\end{document}